\pdfoutput=1
\documentclass[preprint,12pt,authoryear]{elsarticle}

\usepackage{amsmath,amssymb,amsthm}
\usepackage{bm}
\usepackage{booktabs}
\usepackage{array}
\usepackage{graphicx}
\usepackage[hidelinks]{hyperref}
\newif\ifanonymous
\anonymousfalse
\hypersetup{
  pdftitle={Permutation tests for Poisson network regression: Double semi-partialling, studentized statistics and actor effects},
  pdfauthor={David Dekker, Robert W. Krause}
}
\ifanonymous\hypersetup{pdfauthor={}}\fi
\newcommand{\tabnote}[1]{\par\vspace{0.6ex}\begin{minipage}{\linewidth}\footnotesize\textit{Notes.} #1\end{minipage}}
\newtheorem{proposition}{Proposition}
\newtheorem{corollary}{Corollary}
\theoremstyle{remark}
\newtheorem{remark}{Remark}

\newcommand{\E}{\mathbb{E}}
\newcommand{\Var}{\operatorname{Var}}
\newcommand{\Cov}{\operatorname{Cov}}
\newcommand{\tX}{\tilde{X}}
\newcommand{\bZ}{\bm{Z}}
\newcommand{\bW}{\bm{W}}
\newcommand{\bP}{\bm{P}}
\newcommand{\bI}{\bm{I}}
\newcommand{\bg}{\bm{\gamma}}
\newcommand{\trans}{^{\!\top}}

\journal{Social Networks}

\begin{document}

\begin{frontmatter}

\title{Permutation tests for Poisson network regression:\\
Double semi-partialling, studentized statistics and actor effects}

\ifanonymous\else
\author[hw,ufl]{David Dekker\corref{cor1}}
\ead{d.dekker@hw.ac.uk}
\cortext[cor1]{Corresponding author.}
\author[gro]{Robert W. Krause}

\address[hw]{Edinburgh Business School, School of Social Sciences, Heriot-Watt University, Edinburgh, U.K.}
\address[ufl]{Bureau for Economic and Business Research, University of Florida, U.S.}
\address[gro]{National Institute of Education, Nanyang Technological University, Singapore}
\fi

\begin{abstract}
The multiple regression quadratic assignment procedure (MRQAP) is a standard
framework for inference on dyadic network data, and its properties are well
established for the linear model. Network outcomes are often counts, modelled by
Poisson regression with a log link. Double semi-partialling (DSP) is a reparametrization of the fit on the raw
covariates, which holds under any link and which we prove for the log link. The
DSP focal coefficient and its model-based, HC3 and dyadic cluster-robust standard
errors therefore coincide with those of that fit. The choice between relabelling the raw focal array
($X$-permutation) and relabelling its DSP residual acts only on the reference
distribution. The test holds its level when the relabelled designs reproduce the
dependence that the statistic retains, and DSP reproduces the autocorrelation of
the partialled focal covariate. When the raw focal array and its residual share
their autocorrelation, the relabelling test holds the nominal level for all three
statistics, at covariate correlations up to $0.9$ and in networks of $20$ to $100$
actors. A normal reference rejects a true null at rates up to $0.96$. With sender
and receiver effects modelled, every statistic rejects a true null at $0.035$ to
$0.070$ in all conditions, and power against a focal effect of $0.10$ rises from
at most $32\%$ to at least $99.9\%$. When the actor effects cannot be modelled,
adding the sender and receiver means of the nuisance covariates keeps DSP at its
level. We give recommendations and illustrate them on the EIES message-count
network.
\end{abstract}

\begin{keyword}
MRQAP \sep permutation tests \sep Poisson regression \sep
double semi-partialling \sep dyadic data \sep cluster-robust variance
\end{keyword}

\end{frontmatter}

\section{Introduction}\label{sec:intro}

Permutation methods are a default inferential framework for network
regression with dyadic data. Since \citet{mantel1967} and
\citet{hubert1976}, quadratic assignment procedures (QAP) have provided
distribution-free tests for association between relational arrays. These
procedures were extended to multiple regression by \citet{smouse1986}, adapted
for network data by \citet{krackhardt1987,krackhardt1988}, and systematized
for partial tests by \citet{dekker2003,dekker2007}. The core operation is the joint
relabelling of actor rows and columns. The various procedures that go by the
name QAP differ in the restrictions they add to this operation, particularly
in which array is relabelled and what is held fixed. The relabelling itself is
what makes the reference distribution exact under the defining feature of
dyadic data, which is that observations sharing a sender or a receiver are not
independent \citep{warner1979,kenny1984,holland1981,snijders1999}.

Because sender and receiver heterogeneity travels with the actor labels, every
relabelled array carries the same row and column dependence as the observed
network. No parametric adjustment supplies this property directly. For this
reason, permutation remains the natural mode of inference for dyadic data,
where the independence assumptions underlying standard Student-$t$ or normal
reference distributions are violated by construction.

An exact reference operation does not by itself guarantee a valid test. A
permutation test attains its nominal significance level when the statistic
evaluated against the permutation distribution is \emph{pivotal} under the
null hypothesis, meaning that its null distribution is free of nuisance
parameters \citep{good2005,commenges2003}. The question this paper asks is
therefore not whether to permute, but which test statistic can be permuted
safely when the outcome is a count. For the linear model, \citet{dekker2007}
showed that non-pivotal statistics such as raw regression coefficients make
MRQAP tests sensitive to collinearity among dyadic covariates, whereas
studentized statistics largely restore robustness. They also introduced double
semi-partialling (DSP), which residualizes the focal covariate on the nuisance
covariates, keeps the nuisance covariates raw in the outcome model, and
permutes the residual. This approach respects the ancillarity of the nuisance
association \citep[cf.][]{freedman1983,terbraak1992,anderson1999,anderson2001}. 
For a count outcome none of the standard studentized statistics is pivotal.
A second condition then matters as much as the first, namely how much of the
nuisance dependence is approximated in the reference distribution. A
relabelling holds the outcome, the nuisance covariates, and the autocorrelation
of the relabelled array fixed. A statistic whose null distribution depends on
these features alone can still be compared with such a reference. The test is
valid to the extent that the relabelled designs reproduce the
dependence that the statistic retains. What they reproduce depends on whether the raw focal array ($X$-permutation) or its residual (DSP) is relabelled.

Network outcomes such as interaction frequencies, message volumes, citations, and trade flows are counts. The Poisson log-linear model is
the natural specification for such data \citep{mccullagh1989}. It is
consistent under correct specification of the conditional mean alone
\citep{gourieroux1984}, and in the dyadic context it is recommended even when
a log-linearized OLS model appears adequate \citep{santossilva2006}. Count
outcomes have been treated within QAP before, most directly by
\citet{marchi2024}, whose hurdle specification combines actor relabelling with
a raw-coefficient statistic in a design where covariates are uncorrelated. That
orthogonal configuration is the one case where an unstudentized statistic is
safe. \citet{shi2025} recently completed the asymptotic theory of linear
MRQAP. They prove that properly studentized statistics control the Type~I
error rate for partial coefficients across all permutation strategies, whereas
unstudentized statistics do not. Their discussion poses two extensions as open
problems, namely generalizing these results to generalized linear models (GLMs)
and accommodating multiple separate networks permuted independently. This
paper addresses both extensions. Our findings tie the validity of the
permutation test to the autocorrelation that the relabelled array carries, and
its power to the modelling of actor heterogeneity.

Our main result (Section~\ref{sec:res-main}) concerns what the relabelling reference
conditions on. Each relabelled array keeps the row and column autocorrelation of
the array it is drawn from. When the focal covariate and its residual on the
nuisance covariates share that autocorrelation, the test holds its nominal level
for the model-based, HC3 and dyadic cluster-robust statistics alike. It does so
with or without actor effects in the outcome model, at high correlation between
the covariates, and in networks of $20$ to $100$ actors. A normal reference fails in the same designs, because
the dependence between dyads that share an actor inflates the variance of the
focal score far beyond its model-based value. When the collinearity runs through
components with different autocorrelation, the autocorrelation of the residual
differs from that of the raw focal array, and the choice between
$X$-permutation and DSP matters. With $20$ actors per
network DSP then stays close to the nominal level, where $X$-permutation is
conservative in one condition and liberal in the other. The dyadic statistic
depends least on this structure. Without actor effects these errors grow with
the size of the network, also for DSP when the collinearity sits at the actor level. Modelling the sender and receiver heterogeneity
keeps every statistic close to its level in all conditions and restores power. When the actor effects cannot be modelled, adding the
sender and receiver means of the nuisance covariates keeps DSP at its level.

We make five contributions. First
(Section~\ref{sec:fwl}), we show that DSP does not rely on the
Frisch--Waugh--Lovell (FWL) theorem \citep{frisch1933,lovell1963}, which does not
hold under a nonlinear link. The DSP fit is a reparametrization of the raw-$X$
fit. Hence, under a Poisson fixed-effects fit with pooled OLS
residualization, the DSP focal coefficient and its standard error are
\emph{numerically identical} to those of the raw-$X$ fit. Consequently, DSP
and $X$-permutation yield the exact same Wald statistic on the observed data
and differ only through the permutation reference distribution.

Second (Section~\ref{sec:nonpivot}), we show why the model-based and HC3 variance
estimators leave the Wald statistic dependent on the nuisance configuration
under a Poisson link. Because the Poisson Fisher information weights dyads by
their fitted means $\hat\mu$, the effective collinearity is a
$\hat\mu$-weighted multiple correlation. The magnitude of the nuisance effect
governs the dispersion of $\hat\mu$ and thereby acts as a \emph{heterogeneity
amplifier}. The same weighting multiplies the effect of any dependence that the
variance estimator ignores. In our main simulation arms, doubling the nuisance effect raises the inflation of the score variance several times over.
We decompose the resulting failure under a normal reference into a
\emph{scale} component, where the standard error understates the sampling
variability, and a \emph{shape} component, where the studentized statistic is
non-normal in its tails. Scale accounts for nearly all of it.

Third (Section~\ref{sec:reference}), we quantify what the permutation reference
contributes relative to a parametric asymptotic approximation. Because the
exchangeable units are actors rather than dyads, the degrees of freedom for a
Student-$t$ reference are not well defined. We bound the candidate degrees of
freedom between an actor-based and a dyad-based count, and we show that moving within this range hardly changes the rejection rates. Hence, a choice of degrees of freedom cannot repair a statistic whose variance misses or misestimates the dependence. By contrast,
the QAP permutation reference applied to the same statistic on the same fits
reproduces that distortion whenever the relabelled array carries the
autocorrelation of the partialled focal covariate. Permutation is therefore
required for valid inference in this setting.

Fourth (Section~\ref{sec:cluster}), we evaluate the two-way dyadic cluster-robust
variance estimator \citep{fafchamps2007,aronow2015,cameron2011} as a
principled repair for the scale component. It removes the part of the scale
failure that comes from ignoring the dependence between dyads that share an
actor, but two qualifications apply. At realistic actor counts the estimate can
be zero or negative \citep[cf.][]{tabordmeehan2019,cameron2015}, so a rule is
needed for those cases. We exclude such relabellings from the reference
distribution, report their number, and do not compute the test when the
observed estimate is zero or negative. The estimate is also built from one score total per actor, and with skewed
means a few actors contribute most of the score. In our smallest networks it falls well short of the sampling variance. The
relabelling reference reproduces most of that shortfall, which makes the dyadic
statistic the least sensitive of the three to how the collinearity is
structured.

Fifth (Section~\ref{sec:res-largern}), we identify a condition under which neither the
dyadic variance nor DSP is sufficient, and we give a remedy. When the
collinearity runs through components with different autocorrelation and the actor
effects are not modelled, the residual stays correlated with the expected counts
across actors. This occurs, for example, when a covariate is computed
from an actor attribute, such as the difference in citations between two
researchers, because all dyads of an actor then share that attribute. No relabelling keeps this correlation, and the resulting error
grows with the size of the network. Adding the sender mean and the receiver mean
of the nuisance covariates to the outcome model and to the residualization
removes the correlation, and DSP then holds its level.

Throughout the paper, theory and simulation are developed jointly. Every
algebraic identity is verified numerically, and every failure mechanism is
tested against a targeted simulation contrast. Section~\ref{sec:example}
illustrates the argument on the Freeman EIES message-count network
\citep{freeman1979,freeman1980,wasserman1994}, and Section~\ref{sec:disc}
collects practical recommendations.

\section{Background}\label{sec:background}

\subsection{MRQAP and permutation families}

Let $Y$ be a dyadic outcome array on $n$ actors (with the diagonal excluded),
possibly observed across $G$ separate networks, and let $X_1, \bm X_{-1}$ be
dyadic covariates. The QAP operator applies a permutation $\pi$ of
actor labels to rows and columns jointly, mapping dyad $(i,j)$ to
$(\pi(i),\pi(j))$ with $\pi$ drawn independently for each network. Because any sender or
receiver heterogeneity travels with the actor labels, this operation preserves
the row and column dependence structure of the relabelled array. This property
is the source of the procedure's robustness to autocorrelation
\citep{krackhardt1988,dekker2007}.

Procedures that build on this operator differ in the restrictions they add to
accommodate a \emph{partial} test of $H_0\!: \beta_1 = 0$ when the nuisance
parameters $\bg$ are unrestricted. In the terminology of \citet{shi2025}, this
is the \emph{weak} null hypothesis of a zero partial coefficient, as distinct
from the \emph{strong} null of full independence between the arrays. The
distinction matters because a permutation scheme can be exact for the strong
null yet only asymptotically valid for the weak null, which is the hypothesis
applied researchers almost always intend to test. Relabelling the raw focal
array $X_1$ (``$X$-permutation'') breaks the $X_1$--$Y$ association, but it
also alters the observed $X_1$--$\bm X_{-1}$ collinearity in the reference
draws. Residual approaches add an orthogonalization step to avoid this
distortion. Specifically, DSP \citep{dekker2007} residualizes $X_1$ on
$\bm X_{-1}$ by ordinary least squares (OLS), keeps the nuisance covariates
raw in the outcome model, and relabels the residual. This ensures that the
reference distribution is generated at the observed nuisance configuration
\citep[cf.][]{freedman1983,anderson1999}.

Both permutation schemes have a longer history outside network analysis than
their names in this literature suggest. $X$-permutation goes back to \citet{draper1966}, while residualizing the focal covariate on
the nuisance block and permuting that residual was proposed for linear models
by \citet{collins1987} and extended to dyadic data by \citet{dekker2003, dekker2007}. The
residualized procedure goes by different names in different literatures \citep[see][]{terbraak2022, winkler2014}. Here we retain the terms DSP (residual permutation) and (raw) $X$-permutation. Outside the linear model, \citet{potter2005} built a permutation test for
logistic regression on the same construction, residualizing the focal
predictor on the other covariates by OLS and permuting that residual inside
the generalized linear model, and found the type I error controlled at sample
sizes where the asymptotic likelihood-ratio test was not. His test is the
non-network analogue of the procedure analysed in Proposition~\ref{prop:fwl}. Our results connect directly to that wider statistical
literature.

\subsection{Pivotality and the matching reference}

A permutation test attains its nominal level exactly when the test statistic
has the same null distribution for the observed data as for every relabelling
of it. A pivotal statistic, whose null distribution is free of nuisance
parameters, secures this under any relabelling scheme
\citep{good2005,commenges2003}. For the linear-normal model, the studentized
coefficient is exactly pivotal, and \citet{dekker2007} document empirically
that pivotal statistics make MRQAP tests robust to collinearity where
raw-coefficient statistics fail. \citet{anderson1999} and
\citet{andersonterbraak2003} report a complementary finding for residual
methods in ordinary regression and factorial designs, showing that the choice
of permutation scheme matters less than the pivotality of the statistic.
For independent observations, \citet{diciccio2017} prove that permuting a
covariate with correlated covariates present is exact only under the stronger
null of independence, and that under the weak null of a zero coefficient a
statistic studentized by a heteroskedasticity-robust variance is needed for
asymptotic validity. \citet{shi2025} have now made this precise for the linear
dyadic model.
Without studentization, no permutation strategy controls the asymptotic Type~I
error under the weak null, whether one permutes the outcome, the raw focal
covariate, the covariate residual, or the outcome residual. With proper
studentization, all of them control Type~I error asymptotically, and permuting
the raw focal covariate is also finite-sample exact under the strong null.
 
Pivotality is sufficient for a valid test, but it is not necessary. A relabelling
holds a great deal of the nuisance structure fixed. The outcome, the nuisance
covariates, the values of the relabelled array and its row and column
autocorrelation are the same in every draw. A statistic whose null distribution
depends on these features only is compared with relabelled statistics that
depend on them in the same way, and the test holds its level although the
statistic is not pivotal. The statistic has to be free only of what the
relabelling does not hold fixed, which is how the values of the relabelled
array are placed relative to the other variables. Two routes to a valid test
follow, and this paper uses both. One brings the statistic closer to pivotality,
through the variance estimator and through modelling the actor heterogeneity.
The other chooses the relabelled array so that the reference carries the
nuisance dependence the statistic retains. In the Poisson model the standard
studentized statistics are far from pivotal, so the second route carries most
of the weight, and the question becomes how closely the relabelled designs
reproduce the dependence of the observed one. The generalized linear 
model literature has reached the same place by another
route. \citet{hemerik2020} show that nuisance parameters and a misspecified
variance, overdispersion in a Poisson model among them, break the standard
permutation tests for these models, and repair the test by sign-flipping score
contributions after orthogonalizing the focal score against the nuisance
scores. Their observations are independent. The relabelling studied here is the
dyadic counterpart, and the question of what the reference has to reproduce is
the one their orthogonalization answers for independent data.

One result from the linear permutation literature bears directly on the
Poisson case, because maximum likelihood estimation under a log link proceeds
by iteratively reweighted least squares (IRLS) and is therefore a
\emph{weighted} regression. \citet{terbraak2022} shows that the equivalences
that hold in unweighted regression break down once observation weights enter.
Relabelling the predictor and relabelling the response, which are
interchangeable for an overall null test, give different tests in weighted
regression and in partial tests. He compares unweighted versions of each
scheme, which permute the residuals directly, with standardized versions. The standardized versions are more powerful
when the error variance is inversely proportional to the weights, but they can
become severely liberal otherwise. Unweighted predictor permutation, by
contrast, is the most robust to variability in the weights and to the
relationship between weights and error variance, and is recommended for
general use. The DSP procedure we study corresponds to unweighted predictor
permutation in that taxonomy, because the OLS residualization in
\eqref{eq:resid} is unweighted and the resulting residual array is relabelled
directly. Note that in a generalized linear model the weights are estimated rather than
known in advance.

\subsection{Dyadic dependence and robust variances}

Dyads sharing an actor are correlated whenever actors differ systematically in
activity or popularity. This is the dependence captured by the sender and receiver 
variance components of the social relations model \citep{warner1979,kenny1984,snijders1999}, 
by the $p_1$ family \citep{holland1981}, and by the additive and multiplicative
effects models of \citet{hoff2005,hoff2021}, which model it rather than condition on it.
For regression coefficients, this dependence enters the sampling variance through 
covariances between the score contributions of actor-sharing dyads. 
Heteroskedasticity-consistent (HC) estimators \citep{mackinnon1985} ignore these
cross-dyad covariances, and \citet{lindgren2010} shows that ignoring them materially 
distorts inference in dyadic regression. The importance of the variance estimator 
inside a permuted statistic is also established outside network analysis. \citet{helwig2019}
finds that permutation tests of general linear model coefficients built on an unadjusted 
$F$-ratio have high false-positive rates under heteroskedasticity, whereas tests
built on a heteroskedasticity-robust statistic remain asymptotically valid.

The dyadic cluster-robust estimator of \citet{fafchamps2007}, formalized and
extended to directed dyads and GLMs by \citet{aronow2015}, sums score
cross-products over all dyad pairs sharing at least one actor, which
corresponds to two-way clustering on senders and receivers \citep{cameron2011}.
\citet{marrs2023} derive the variance of a regression coefficient under
exchangeable relational errors, show that it has the same actor-sharing
structure, and give an estimator that is unbiased in finite samples where the
dyadic cluster-robust estimator is biased downward when the actors are few.
The studentization that \citet{shi2025} prove valid for symmetric
single-network MRQAP belongs to the same family. Their variance estimator is
the one-way cluster-robust form of \citet{liang1986} scaled by a factor of
four, which is the undirected specialization of the dyadic-robust construction
used here. Its finite-sample reliability, however, is governed by the number
of \emph{actors} ($n$) rather than the number of dyads ($N$).
\citet{tabordmeehan2019} establishes conditions for asymptotic normality of
the dyadic-robust $t$-statistic along with a finite-sample correction, and the
broader clustering literature warns that having few clusters makes such
estimators noisy and non-normal \citep{cameron2015}. Hence, in a network
with few actors the dyadic cluster-robust estimate can be expected to be noisy and
too small. Furthermore, the estimate is not guaranteed to be positive
\citep[cf.][]{cameron2011}. When it is zero or negative, the studentized statistic
does not exist.

\section{Theory}\label{sec:theory}

\subsection{Model and estimators}\label{sec:model}

Index directed dyads by $\ell = 1,\dots,N$, where each dyad $\ell$ is an
ordered pair $(i,j)$ with $i \ne j$ within network $g(\ell) \in \{1,\dots,G\}$.
The count outcome follows the Poisson specification
\begin{equation}
  Y_\ell \mid \bm X \sim \mathrm{Poisson}(\mu_\ell), \qquad
  \log \mu_\ell = \beta_1 X_{1,\ell} + \bm x_{-1,\ell}\trans \bg + \alpha_{g(\ell)},
  \label{eq:model}
\end{equation}
with focal covariate $X_1$, nuisance covariates $\bm x_{-1}$, and network
fixed effects $\alpha_g$. Collect the nuisance covariates and fixed-effect
indicators in the $N \times q$ matrix $\bZ$ and assume that $[X_1, \bZ]$ has
full column rank. Let $\bP_{\bZ} = \bZ(\bZ\trans\bZ)^{-1}\bZ\trans$ denote the
unweighted OLS projector onto $\operatorname{col}(\bZ)$. Structural row and
column autocorrelation, in the sense of \citet{dekker2003}, is induced by
latent sender and receiver random effects $u_i$ and $v_j$ added to the linear
predictor, with a common standard deviation $\sigma_R$ set in
Section~\ref{sec:design}.

The DSP construction residualizes only the focal covariate via OLS,
\begin{equation}
  \tX_1 = (\bI - \bP_{\bZ})\, X_1,
  \label{eq:resid}
\end{equation}
fits \eqref{eq:model} on $[\tX_1, \bZ]$ with the nuisance block entering raw,
and permutes $\tX_1$ by QAP. We use the term ``pooled'' for a single
residualization across the stacked networks, with network means removed via
$\bZ$. The alternative ``per-network'' residualization fits a separate nuisance
slope within each network and is discussed in Remark~\ref{rem:scope}. Let
$\hat\beta_1^{X}$ and $\hat\beta_1^{\mathrm{DSP}}$ denote the focal
maximum-likelihood coefficients from the fits on $[X_1,\bZ]$ and $[\tX_1,\bZ]$,
respectively. The studentized Wald statistic under a given variance estimator
is $t = \hat\beta_1 / \widehat{\mathrm{se}}(\hat\beta_1)$.

\subsection{DSP as a reparametrization of the raw fit}\label{sec:fwl}

In the linear model the properties of DSP are usually derived from the
Frisch--Waugh--Lovell (FWL) theorem \citep{frisch1933,lovell1963}. The theorem
states that the coefficient of the focal covariate in the full regression equals
the coefficient in the regression of the residualized outcome on the residualized
focal covariate alone, with the nuisance block dropped. \citet{ding2021} shows
that this equivalence extends to several standard errors. The theorem rests on
the orthogonal geometry of least squares. It does not carry over to
maximum-likelihood fitting under a log link, where residualizing the count
outcome has no exact counterpart.

However, DSP does not need the theorem. DSP retains the nuisance covariates in
the outcome model by construction and only replaces the focal covariate by its
OLS residual on the nuisance block. The designs $[X_1,\bZ]$ and $[\tX_1,\bZ]$
span the same column space. Hence, the DSP fit is a non-degenerate
reparametrization of the raw fit. This argument uses no property of least
squares. It applies to any model in which the covariates enter through the
linear predictor alone, and hence under any link. We prove it for the Poisson
model with log link, where the log-likelihood is strictly concave and the
maximizer is therefore unique. When the log-likelihood is not concave, for example in a negative binomial model with estimated dispersion, the solutions of the two fits still correspond one to one, but uniqueness is not guaranteed. Retaining the nuisance block matters also in the
linear model. \citet{ding2021} shows that regressing the outcome on the
residualized covariate alone gives the same coefficient but not the same
standard errors, because the residuals differ from those of the full regression.
DSP avoids this, since its residuals are those of the full fit.
Proposition~\ref{prop:fwl} states the result.

\begin{proposition}[Reparametrization invariance]\label{prop:fwl}
Consider the Poisson ML fits of \eqref{eq:model} on $[X_1,\bZ]$ and on
$[\tX_1,\bZ]$, with $\tX_1$ defined as in \eqref{eq:resid} and the nuisance
block entering raw. The fitted linear predictors coincide, and
$\hat\beta_1^{\mathrm{DSP}} = \hat\beta_1^{X}$ holds exactly for any value of
the nuisance coefficients and any degree of structural heterogeneity.
\end{proposition}

\begin{corollary}\label{cor:se}
The model-based, HC3, and dyadic cluster-robust standard errors also coincide
across the two fits, meaning that
$\widehat{\mathrm{se}}(\hat\beta_1^{\mathrm{DSP}}) =
\widehat{\mathrm{se}}(\hat\beta_1^{X})$. Consequently, DSP and the raw-$X$
partial fit yield the \emph{exact same} Wald statistic on the observed data,
and can differ only through the permutation reference distribution.
\end{corollary}

Proofs are provided in \ref{app:proofs}. All three variances are computed
from the full fit, and the leverages in HC3 are those of the full weighted design. Under the fixed-effects fit and under
the penalized fit with actor effects alike, residualization produces the exact
same point estimate and Wald statistic as the raw covariate fit. Any difference
between $X$-permutation and DSP therefore lies entirely in the reference
distributions they generate. Under $X$-permutation, each draw relabels the raw
focal array while the nuisance block stays fixed. Averaged over all
relabellings, the relabelled array is uncorrelated with the nuisance block, so its partialled version is essentially the raw focal array, and the reference distribution reflects the row and column autocorrelation of that array. The observed statistic depends instead on the focal covariate after
the nuisance block has been partialled out, whose autocorrelation can differ from that of the raw focal array. Under DSP, the relabelled
array is the residual itself, so the reference reflects the autocorrelation of
the partialled focal covariate. Two features of the observed design are lost
under either scheme. The fitted-mean weighting in \eqref{eq:vif} below leaves a
small weighted correlation between $\tX_1$ and $\bZ$, and a relabelling of the
focal array alone cannot keep any correlation between it and the nuisance
covariates at dyads that share an actor.

Two implications follow from this equivalence. First, comparing the two
permutation schemes becomes a clean one-factor contrast, because the model
fit, point estimate, standard error, and observed data are common to both and
only the reference distribution differs. Second, evaluating the benefit of
residualization becomes a quantitative empirical question rather than a
qualitative theoretical one. Because theory fixes the observed statistic
without bounding the finite-sample behaviour of the reference draws under a
log link, the practical contribution of DSP must be evaluated directly from the simulated rejection rates.

\begin{remark}[Scope conditions]\label{rem:scope}
Two conditions are essential to Proposition~\ref{prop:fwl}. (i)~\emph{The
residualization space.} The OLS residualization must remove a subspace of the
fixed-effect nuisance design $\operatorname{col}(\bZ)$ included in the outcome
model. Removing a smaller subspace leaves the algebraic identity intact, since
the removed component still lies in $\operatorname{col}(\bZ)$, but leaves
residual collinearity in the relabelled array. Removing more than
$\operatorname{col}(\bZ)$ alters the column space and changes the estimator.
For example, per-network residualization under a common-slope outcome model
removes network-specific nuisance slopes not present in $\bZ$, which breaks
the identity. Per-network residualization is appropriate only when the outcome
model itself includes network-specific nuisance slopes. Under random-effects
fitting, the same principle applies with one addition, namely that any
random-effect design matrices for networks or actors must be excluded from the
OLS residualization. Residualizing on random-effect indicators reparameterizes
the penalized coordinates and changes the penalized likelihood objective, so
the two fits no longer share a maximizer. This holds for any permutation scheme
and is not specific to QAP. Residualizing on the fixed-effect
design $\bZ$ alone leaves the penalty untouched, and the identity holds at the
converged maximizer. (ii)~\emph{Converged fits.} The column-space argument
applies at the exact maximum-likelihood solution. An iterative optimizer that
stopped short of convergence leaves the two fits at different points along the
likelihood surface. Because the raw design is collinear whereas the
residualized design is orthogonalized, a fixed iteration budget carries the
two fits to different distances from the maximizer. This numerical gap is easy
to mistake for a substantive property of the method. Apparent
failures of the identity in practice are diagnostics of the
residualization design or of incomplete optimizer convergence, rather than
departures from the theory, but they can also arise from different missing data 
treatments based on partially observed covariates. In this study, we assume all
data are observed for all dyads.
\end{remark}

\subsection{How the nuisance effect enters the Wald statistic}\label{sec:nonpivot}

In a Poisson model with log link the Fisher information of the coefficients is
the cross-product of the design matrix weighted by
$\bW = \operatorname{diag}(\hat{\bm\mu})$ \citep{mccullagh1989}. Each dyad thus
counts in proportion to its fitted mean, and we refer to projections with these
weights as projections in the $\hat\mu$-weighted metric. After partialling out the
nuisance block in this metric, the effective focal information is
\begin{equation}
  \mathcal I_{11\cdot 2}
  = \bigl(X_1\trans \bW X_1\bigr)\bigl(1 - R^2_{1\cdot 2}\bigr),
  \qquad
  R^2_{1\cdot 2}
  = \frac{X_1\trans \bW \bZ (\bZ\trans \bW \bZ)^{-1} \bZ\trans \bW X_1}
         {X_1\trans \bW X_1},
  \label{eq:vif}
\end{equation}
The model-based variance of
$\hat\beta_1$ is $\mathcal I_{11\cdot2}^{-1}$, and the factor
$(1 - R^2_{1\cdot2})^{-1}$ is the $\hat\mu$-weighted variance inflation factor
(VIF). This quantity is the exact analogue of the linear variance inflation
factor, with the critical difference that the projection is evaluated under
the fitted-mean weights $\bW$. Two consequences follow.

\emph{The nuisance effect acts as a heterogeneity amplifier.} Through the
weight matrix $\bW$, the magnitude of the nuisance coefficient $\bg$ governs
the dispersion of the fitted means, as $\mathrm{sd}(\log\hat\mu)$ grows with
$\|\bg\|$. This dispersion concentrates Fisher information onto high-$\hat\mu$
dyads and raises $R^2_{1\cdot2}$ relative to the unweighted correlation. It also
multiplies the effect of any dependence that the variance estimator ignores,
because the same high-$\hat\mu$ dyads dominate the variance of the score. We will quantify this inflation for our designs as the ratio $\kappa$ of the variance of the score to its model-based value. The simulation results are organized around $\kappa$ because it measures what the relabelled designs have to reproduce. The $\hat\mu$-weighted VIF itself is reported for the empirical illustration. This
mechanism has no counterpart in the linear model, where $\bW = \sigma^2\bI$
regardless of $\bg$. This coupling is also where the linear-model guarantee of
\citet{shi2025} fails to transfer. While studentization secures asymptotic
weak-null validity for every permutation strategy in the linear model, under
the log link the null distribution of the studentized statistic depends on the
nuisance configuration through $\bW$, and standard studentization cannot remove
that dependence.

\emph{Scale and shape distortions.} The studentized statistic $T$ can depart
from standard normality in two separable ways, which we distinguish using a
first-order decomposition \citep[cf.][]{hall1992}. Let $s$ be the robust
empirical scale of the null distribution of $T$, estimated across null
replications as $\operatorname{median}(|T|)$ divided by
$\Phi^{-1}(0.75) \approx 0.6745$, so that $s = 1$ for a standard normal
statistic. Using a median-based scale estimate is deliberate, because the
moment-based ratio of the empirical variance of $\hat\beta_1$ to the mean
estimated variance $\E[\widehat{\mathrm{se}}^2]$ is sensitive to a small
number of extreme fits and says little about the bulk of the
distribution. If the only departure from normality were scale inflation
($s > 1$ for an otherwise Gaussian statistic), a test referring $T$ to
$\mathcal N(0,1)$ at nominal level $\alpha = 0.05$ would reject at rate
$2\Phi(-1.96/s)$. The difference between this scale-only prediction and the
rejection rate actually observed under the normal reference isolates the
residual \emph{shape} or tail component. Because this decomposition is a property of the observed 
statistic $T$, Corollary~\ref{cor:se} guarantees
that it is identical under $X$-permutation and under DSP.
What differs between the two permutation schemes is only how much of the
excess their reference distributions reproduce. Separating scale from shape
matters because a cluster-robust variance estimator repairs scale
understatement directly, whereas the permutation reference accounts for scale
and shape distortions only insofar as actor relabelling reproduces them across
draws.

\subsection{The reference distribution}\label{sec:reference}

We use permutation inference because the structural dependencies of dyadic
data violate the assumptions required by parametric reference distributions.
Because dyads sharing an actor are dependent, the exchangeable sampling unit
is the actor rather than the dyad, and the degrees of freedom $\nu$ for a
Student-$t$ reference are not identified. A count based on the dyads gives
$\nu = N - k$, with $k$ the number of coefficients. A count based on the actors
gives a much smaller value.

However, the choice within this range has little effect. The Student-$t$
quantiles are close to the normal quantiles already for about twenty degrees of
freedom, so the reference hardly changes with $\nu$. Widening the tails slightly
cannot calibrate a statistic whose null scale differs from one by the factors of
Section~\ref{sec:nonpivot}. Hence, the failure of parametric tables is one of
scale and shape rather than of tail thickness (\ref{app:reference} gives the
numerical comparison for our design).

The QAP permutation reference, by contrast, is generated from the observed
data using the exact same fitting procedure. Each relabelled statistic is computed on a design with the same actor
structure and studentized by the same estimator, so the reference reproduces
distortions that a fixed table cannot. It reproduces them only to the extent
that the relabelled designs share the features of the observed design on which
the null distribution depends.

\subsection{The dyadic cluster-robust variance and non-positive estimates}\label{sec:cluster}

Scale inflation in standard Wald statistics originates in variance estimators
that ignore cross-dyad score covariances. The two-way dyadic cluster-robust
estimator \citep{fafchamps2007,aronow2015} replaces the diagonal HC ``meat''
$\sum_\ell s_\ell^2$, where $s_\ell = (Y_\ell - \hat\mu_\ell)\,
x^{\perp}_{1,\ell}$ is the partialled focal score contribution, with the
double sum
\begin{equation}
  M \;=\; \sum_{\ell=1}^N\sum_{m=1}^N s_\ell\, s_m\,
  \mathbf 1\{\text{dyads } \ell \text{ and } m \text{ share at least one actor}\}.
  \label{eq:meat}
\end{equation}
This quantity is computed as a two-way clustering on senders and receivers
with an overlap correction \citep{cameron2011}, and the resulting focal
variance estimate is $\hat V_{\mathrm{dyad}} = \mathcal I_{11\cdot2}^{-1} M\,
\mathcal I_{11\cdot2}^{-1}$.

While $\hat V_{\mathrm{dyad}}$ targets the correct sampling variance under
actor-induced dependence, its plug-in estimate converges in the number of
actors $n$ rather than the number of dyads $N$. Furthermore, the indicator
kernel in \eqref{eq:meat} is not positive semidefinite, so $M$ is a quadratic form that is not
guaranteed to be positive in finite samples. In the two-way clustering
computation this appears as the subtracted overlap term. When $M$ is zero or negative, the
variance estimate does not exist, and neither does the studentized statistic.

We apply the same rule to the observed fit and to every relabelled fit. If under a relabelling the dyadic variance estimate
is zero or negative, that relabelling is excluded from the reference distribution, and the
number of excluded relabellings is reported. When the observed estimate is
zero or negative, the dyadic test is not computed for that data set. Under
exact exchangeability this rule keeps the test valid, because applying the
same exclusion to the observed and the relabelled statistics leaves the
retained ones exchangeable.

The dyadic variance removes the part of the scale failure that comes from
ignoring the dependence between dyads that share an actor, but it does not
make the statistic pivotal. With few actors the estimate is noisy and falls short of the sampling variance of $\hat\beta_1$, and the null scale of the statistic remains above one. The relabelled statistics share this excess. Each
is studentized by the same estimator, on a design with the same actor
structure, so the reference distribution is wider than the standard normal by
nearly the same factor. How close the match is, with and without actor effects
in the model and when the focal covariate and its residual differ in
autocorrelation, is a question for the simulations.

\section{Simulation design}\label{sec:design}

\paragraph{Networks}
Each simulated data set consists of $G = 3$ independent groups of $n = 20$
actors. Within a group every ordered pair of distinct actors is a dyad, so a
group has $n(n-1) = 380$ dyads and a data set has $N = 1{,}140$. Each dyad
carries one value of the outcome $Y$ and of the two covariates $X_1$ and $X_2$,
so a data set holds three directed networks for each of the three variables.
One Poisson regression is fitted to all $N$ dyads, with a separate intercept
for each group \eqref{eq:ygen}. No actor appears in more than one group, and
every actor-level quantity below is drawn separately for each group, which
makes the groups independent given the covariates. A relabelling permutes the
actors of each group independently and applies the same permutation to the
rows and the columns of that group.

\begin{figure}[t]
\centering
\includegraphics[width=\textwidth]{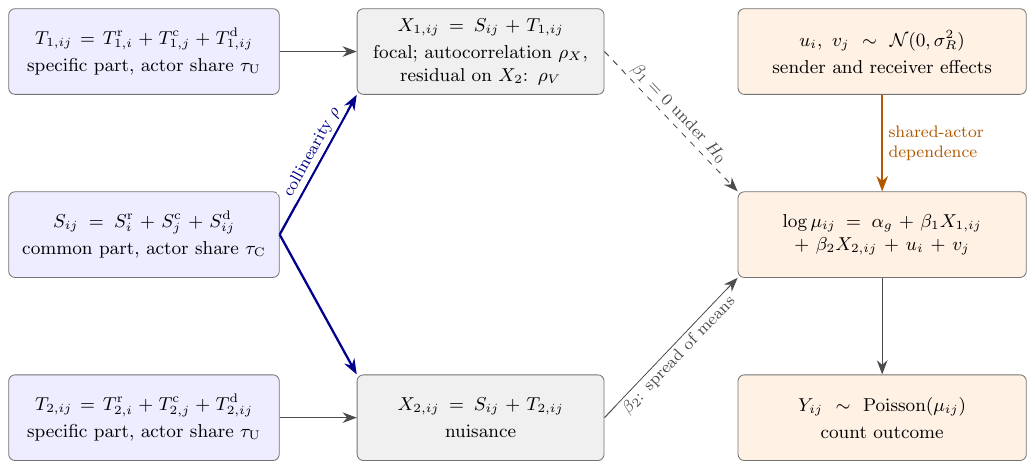}
\caption{Sources of dependence in the simulated data. Each covariate, $X_k$ $(k \in \{1,2\})$, is the sum
of a part $S$ common to $X_1$ and $X_2$ and a specific part $T_k$, and every
part has a sender (r), a receiver (c) and a dyad (d) component. The actor shares
$\tau_{\mathrm C}$ and $\tau_{\mathrm U}$ set the row and column autocorrelation of the raw focal array ($\rho_X$) and of its residual on $X_2$ ($\rho_V$).
The common part makes the covariates collinear, the sender and receiver effects
make the errors of dyads that share an actor dependent and overdispersed in a
fit without them, and the nuisance effect $\beta_2$ spreads the means. Under the
null the focal covariate has no effect (dashed).}
\label{fig:dgp}
\end{figure}

\paragraph{Covariates}
Both covariates are counts built from independent Poisson components, following the
row, column and cell decomposition of \citet[eq.~19]{dekker2007}.
Figure~\ref{fig:dgp} shows how the parts of the design generate the dependence
that the tests have to handle. For dyad $(i,j)$
of network $g$ and $k = 1, 2$,
\begin{equation}
  X_{k,ij} \;=\;
  \underbrace{S^{\mathrm r}_{i} + S^{\mathrm c}_{j} + S^{\mathrm d}_{ij}}_{\text{common part } S_{ij}}
  \;+\;
  \underbrace{T^{\mathrm r}_{k,i} + T^{\mathrm c}_{k,j} + T^{\mathrm d}_{k,ij}}_{\text{specific part } T_{k,ij}} ,
  \label{eq:xgen}
\end{equation}
where the superscripts $r$, $c$ and $d$ mark a sender (row), receiver (column) and dyad
(cell) component. The components are mutually independent Poisson variables with
means
\begin{equation}
\begin{aligned}
  \E S^{\mathrm r}_{i} = \E S^{\mathrm c}_{j} &= \rho\,\tau_{\mathrm C}\,\lambda, &
  \E S^{\mathrm d}_{ij} &= \rho\,(1 - 2\tau_{\mathrm C})\,\lambda, \\
  \E T^{\mathrm r}_{k,i} = \E T^{\mathrm c}_{k,j} &= (1-\rho)\,\tau_{\mathrm U}\,\lambda, &
  \E T^{\mathrm d}_{k,ij} &= (1-\rho)\,(1 - 2\tau_{\mathrm U})\,\lambda .
\end{aligned}
\label{eq:compmeans}
\end{equation}
Here $\lambda = 4$ is the mean of each covariate, $\rho$ the correlation between the
two covariates, and $\tau_{\mathrm C}, \tau_{\mathrm U} \in [0, \tfrac12)$ are the
actor shares of the common part and of the specific parts. A sum of independent
Poisson variables is again Poisson, so at every dyad each covariate is Poisson with
mean and variance $\lambda$ (standard deviation 2). The covariance structure follows
from \eqref{eq:xgen} without approximation. At the same dyad,
$\Cov(X_{1,ij}, X_{2,ij}) = \rho\lambda$, so the correlation between the covariates
is $\rho$ exactly. For two dyads that share a sender, $(i,j)$ and $(i,l)$ with
$j \ne l$,
\begin{equation}
  \begin{aligned}
  \Cov(X_{k,ij}, X_{k,il}) &= \rho_X\,\lambda, \qquad
  \rho_X = \rho\,\tau_{\mathrm C} + (1-\rho)\,\tau_{\mathrm U}, \\
  \Cov(X_{1,ij}, X_{2,il}) &= \rho\,\tau_{\mathrm C}\,\lambda ,
  \end{aligned}
  \label{eq:xcov}
\end{equation}
and the same holds for two dyads that share a receiver. All other pairs of dyads are
uncorrelated, including the reciprocal pair $(i,j)$ and $(j,i)$ and pairs in
different networks. Thus $\rho_X$ is the row and column autocorrelation of each
covariate in the sense of \citet{dekker2007}. The grid varies $\rho$ over
$\{0, 0.3, 0.6, 0.7, 0.8, 0.9\}$ and the other parts use $\rho = 0.6$. Because the
correlation is set exactly rather than calibrated, the realized correlation equals
the target up to sampling error.

\paragraph{Structure of the collinearity}
The way the autocorrelation is divided between the common and the specific parts
determines what DSP relabels. The population regression of $X_1$ on $X_2$ has slope
$\rho$, so the residual array of \citet[eq.~3]{dekker2007} is
$V_{ij} = X_{1,ij} - \rho X_{2,ij} - (1-\rho)\lambda$. It has variance
$(1-\rho^2)\lambda$ and, for two dyads that share a sender or a receiver,
\begin{equation}
  \Cov(V_{ij}, V_{il}) = \rho_V \,(1-\rho^2)\,\lambda, \qquad
  \rho_V = \frac{(1-\rho)\,\rho\,\tau_{\mathrm C} + (1+\rho^2)\,\tau_{\mathrm U}}{1+\rho} .
  \label{eq:vcov}
\end{equation}
By construction $V$ is uncorrelated with $X_2$ at the same dyad. At dyads that share
an actor, however, $\Cov(V_{ij}, X_{2,il}) = \rho(1-\rho)(\tau_{\mathrm C} -
\tau_{\mathrm U})\lambda$, which vanishes only when the two shares are equal. The regression of
$X_1$ on $X_2$ has one slope for both levels. When the shares differ, this slope is
too small for the actor-level parts in H1 and too large for them in H2, and the
residual remains correlated with $X_2$ across actors. We use
three conditions (Table~\ref{tab:covstruct}). In the main condition
$\tau_{\mathrm C} = \tau_{\mathrm U} = \rho_a$, so that $\rho_X = \rho_V = \rho_a$
and the residual array has the same autocorrelation as the raw focal array. In
condition H1 the common part is almost entirely actor-level
($\tau_{\mathrm C} = 0.45$) and the specific parts are dyad-level
($\tau_{\mathrm U} = 0$). The collinearity then sits at the actor level, and the
variation in $X_1$ that identifies $\beta_1$ is mostly dyadic. Condition H2 reverses
the roles ($\tau_{\mathrm C} = 0$, $\tau_{\mathrm U} = 0.45$), so the collinearity
sits at the dyad level and the identifying variation is mostly actor-level. An actor
attribute makes the two conditions concrete. A covariate that repeats an attribute
of the sender over all dyads of that sender, $x_{ij} = a_i$, has only actor-level
variation, and the sender and receiver parts of \eqref{eq:xgen} are covariates of
this kind. In H1 the two covariates share such attributes and differ in their
dyad-level parts. Partialling $X_2$ out of $X_1$ then removes most of the
attribute, and the residual is mostly dyadic. In H2 the covariates share a
dyad-level part and each carries attributes of its own. Partialling removes the
shared dyadic variation and leaves the attributes of $X_1$ in the residual. H1 and
H2 leave the variance of each covariate, the correlation $\rho$ between them and the
outcome model unchanged. What they change is the covariance structure of the
residual array relative to that of the raw focal array.

In data, $\rho_X$ and $\rho_V$ are estimated from the arrays themselves. For an array
$x$ on a network of $n$ actors, let $\tilde x_{ij}$ denote its entries centred at the
network mean. The sample row autocorrelation is
\begin{equation}
  \hat\rho_{\mathrm{row}}(x) \;=\;
  \frac{\sum_{i}\sum_{j \ne l} \tilde x_{ij}\,\tilde x_{il}}
       {(n-2)\,\sum_{i}\sum_{j} \tilde x_{ij}^{\,2}} ,
  \label{eq:rhohat}
\end{equation}
where $j$ and $l$ run over the receivers of sender $i$. This is the average product
of two different entries in the same row, divided by the average squared entry. The
sample column autocorrelation $\hat\rho_{\mathrm{col}}(x)$ exchanges the roles of
senders and receivers, and with several networks the sums run over the networks. In our
simulations senders and receivers are generated alike and we report the mean of
the two. In data the two can differ and are reported separately. Applied to the raw focal array, \eqref{eq:rhohat}
estimates $\rho_X$. Applied to the DSP residual, it estimates $\rho_V$.

\begin{table}[tbp]
\centering
\caption{Covariance structure of the covariates at $\rho = 0.6$.}
\label{tab:covstruct}
\begin{tabular}{lccccc}
\toprule
Condition & $\tau_{\mathrm C}$ & $\tau_{\mathrm U}$ & $\rho_X$ & $\rho_V$ &
$\operatorname{corr}(V_{ij}, X_{2,il})$ \\
\midrule
Main ($\rho_a = 0.30$)       & 0.30 & 0.30 & 0.30 & 0.30 & 0 \\
H1, actor-level collinearity & 0.45 & 0    & 0.27 & 0.07 & $\phantom{-}0.14$ \\
H2, dyad-level collinearity  & 0    & 0.45 & 0.18 & 0.38 & $-0.14$ \\
\bottomrule
\end{tabular}
\tabnote{$\rho_X$ is the
row and column autocorrelation of each covariate \eqref{eq:xcov}, $\rho_V$ that of
the residual of $X_1$ on $X_2$ \eqref{eq:vcov}, and the last column the correlation
between the residual and the nuisance covariate at two dyads that share an actor.}
\end{table}

\paragraph{Outcome}
Given the covariates and the actor effects, the counts are independent with
\begin{equation}
  Y_{ij} \mid \bm X, \bm u, \bm v \sim \mathrm{Poisson}(\mu_{ij}), \qquad
  \log \mu_{ij} = \alpha_g + \beta_1 X_{1,ij} + \beta_2 X_{2,ij} + u_i + v_j ,
  \label{eq:ygen}
\end{equation}
with network intercepts $(\alpha_1, \alpha_2, \alpha_3) = (-1.2, -1.0, -0.8)$ and
sender and receiver effects $u_i, v_j \sim \mathcal N(0, \sigma_R^2)$, drawn
independently for every actor of every network. There is no dyad-level error term
in \eqref{eq:ygen}. Dyad-level variation in the outcome comes from the covariates
and from the Poisson draw. We set $\sigma_R^2 = \rho_a / (1 - 2\rho_a)$, which is
the variance that would give a latent variable $u_i + v_j + \varepsilon_{ij}$ with
$\Var(\varepsilon_{ij}) = 1$ a row and column autocorrelation of $\rho_a$
\citep{dekker2007}. For $\rho_a = 0$, $0.15$ and $0.30$ this gives
$\sigma_R = 0$, $0.46$ and $0.87$. Because there is no $\varepsilon_{ij}$, $\rho_a$
indexes the strength of the actor effects rather than the realized autocorrelation of
$Y$. The same $\rho_a$ sets the actor shares of the covariates in the main
condition.

Writing $m_{ij} = \E(Y_{ij} \mid \bm X) = \exp(\alpha_g + \beta_1 X_{1,ij} +
\beta_2 X_{2,ij} + \sigma_R^2)$ and $e_{ij} = Y_{ij} - m_{ij}$ for the error of a
model without actor effects, \eqref{eq:ygen} implies
\begin{equation}
  \begin{aligned}
  \Var(e_{ij} \mid \bm X) &= m_{ij} + m_{ij}^2\bigl(e^{2\sigma_R^2} - 1\bigr), \\
  \Cov(e_{ij}, e_{il} \mid \bm X) &= m_{ij}\, m_{il}\bigl(e^{\sigma_R^2} - 1\bigr)
  \qquad (j \ne l),
  \end{aligned}
  \label{eq:ecov}
\end{equation}
with the same covariance for two dyads that share a receiver, and zero covariance for
every other pair of dyads, including reciprocal ones. The first term of the variance
is the Poisson variance and the second the overdispersion created by the actor
effects. The covariance is the dependence between dyads that share an actor.
Conditional on $\bm u$ and $\bm v$ the errors are independent with variance
$\mu_{ij}$. The covariate structure, and with it the choice among the main
condition, H1 and H2, enters \eqref{eq:ecov} only through the means $m_{ij}$.

\paragraph{How the covariate structure reaches the test}
For a fit without actor effects, the score for $\beta_1$ is
$U = \sum_\ell \tilde x_\ell\, e_\ell$, where $\tilde x$ is the focal covariate
partialled out of the nuisance design in the weighted metric of
\eqref{eq:vif}. Let $\mathcal A$ be the set of ordered pairs of distinct dyads that
share an actor. Given the covariates,
\begin{equation}
  \Var(U \mid \bm X) \;=\;
  \sum_{\ell} \tilde x_\ell^{\,2}\, \Var(e_\ell \mid \bm X)
  \;+\; \bigl(e^{\sigma_R^2} - 1\bigr)
  \sum_{(\ell, m) \in \mathcal A} \tilde x_\ell\, \tilde x_m\, m_\ell\, m_m ,
  \label{eq:uvar}
\end{equation}
whereas the model-based variance uses $\sum_\ell \tilde x_\ell^{\,2} m_\ell$. The
first sum carries the overdispersion, which HC3 estimates, and the second the
dependence between dyads that share an actor, which only the dyadic estimator
\eqref{eq:meat} targets. The second sum depends on how $\tilde x$ covaries across
dyads that share an actor, and to first order that covariance is governed by
$\rho_V$. A relabelling moves the values of an array together with the
actors. Entries that shared a sender or a receiver before the relabelling still
share one afterwards, so the row and column autocorrelation of the relabelled
array is unchanged.
Relabelling the raw focal array therefore produces reference designs whose partialled
focal covariate has autocorrelation $\rho_X$, while relabelling the DSP residual
keeps $\rho_V$. Hence, we expect the two schemes to give the same test when
$\rho_X = \rho_V$. When $\rho_X > \rho_V$ the reference of $X$-permutation should
be too wide and the test conservative, and when $\rho_X < \rho_V$ too narrow and
the test liberal. Table~\ref{tab:kappa} makes this concrete. It reports the ratio
$\kappa$ of \eqref{eq:uvar} to the model-based variance, evaluated at the true
means $m$, for the observed designs and for relabelled ones. In the main condition
both relabellings reproduce $\kappa$ to within about 15\%. In H1 $X$-permutation overstates it and in H2 it understates it, by up to two thirds, whereas DSP stays within 27\%. Given the actor effects, the errors are
independent, so a fit that includes them faces $\kappa = 1$ apart from the error in
the estimated effects, and the distinction disappears.

The second sum also depends on where the large values of $\tilde x$ sit. Its
terms carry the products $\tilde x_\ell m_\ell$, so the sum is larger when actors
with large expected counts also have large values of $\tilde x$, that is, when the
two are correlated across actors. Under relabelling the covariate values move to
other actors and the expected counts do not. Hence, this correlation is zero on
average, whatever its observed value, and we expect the reference of either scheme
to be too narrow when the correlation is present. Furthermore, the number of pairs
in $\mathcal A$ grows with $n^3$ and the number of dyads with $n^2$, so both
effects should grow with the size of the network.

Finally, the dyadic estimator \eqref{eq:meat} estimates the second sum from one
score total per actor. When few actors carry most of the score, we expect it to
fall short of \eqref{eq:uvar} and the dyadic statistic to be liberal under a normal
reference. The relabelled statistics are studentized by the same estimator, so the
relabelling test should hold its level also for this statistic.

\begin{table}[tbp]
\centering
\caption{Inflation $\kappa$ of the variance of the focal score \eqref{eq:uvar}
relative to the model-based variance, for fits without actor effects, at $\rho = 0.6$
and $\rho_a = 0.30$.}
\label{tab:kappa}
\footnotesize
\setlength{\tabcolsep}{2.9pt}
\begin{tabular}{llrrrrrrrr}
\toprule
 & & \multicolumn{4}{c}{Observed} & \multicolumn{2}{c}{Raw $X$ relabelled} & \multicolumn{2}{c}{DSP relabelled} \\
\cmidrule(lr){3-6}\cmidrule(lr){7-8}\cmidrule(lr){9-10}
$\beta_2$ & Condition & $\kappa$ & Actor & \begin{tabular}[b]{@{}r@{}}Eff.\\ actors\end{tabular} & Align. & $\kappa$ & Actor & $\kappa$ & Actor \\
\midrule
0.25 & Main & 46.2  & 32.1  & 17.0 & $.00$  & 43.0 ($-7\%$)   & 30.1  & 42.5 ($-8\%$)   & 29.6 \\
     & H1   & 24.1  & 9.8   & 19.1 & $+.54$ & 40.1 ($+66\%$)  & 27.1  & 19.5 ($-19\%$)  & 6.5 \\
     & H2   & 54.0  & 39.9  & 17.4 & $-.34$ & 29.8 ($-45\%$)  & 16.8  & 50.3 ($-7\%$)   & 37.4 \\
\midrule
0.50 & Main & 332.6 & 168.7 & 9.4 & $.00$  & 283.6 ($-15\%$) & 145.8 & 281.2 ($-15\%$) & 143.1 \\
     & H1   & 238.4 & 62.6  & 8.5 & $+.57$ & 279.6 ($+17\%$) & 132.9 & 173.8 ($-27\%$) & 22.9 \\
     & H2   & 368.1 & 204.0 & 9.9 & $-.38$ & 219.4 ($-40\%$) & 72.8  & 316.2 ($-14\%$) & 174.5 \\
\bottomrule
\end{tabular}
\tabnote{``Observed'' uses the generated design. The next columns use the
same design after one relabelling of the raw focal array or of the DSP residual, with
the change relative to the observed value in parentheses. ``Actor'' is the part of
$\kappa - 1$ that comes from the second sum of \eqref{eq:uvar}, the covariance between
dyads that share an actor. The rest of $\kappa - 1$ is the overdispersion. ``Eff.\
actors'' is the effective number of actors that carry the focal score in the observed
design, $(\sum_i v_i)^2 / \sum_i v_i^2$ with $v_i$ the variance of the part of the
score that runs through actor $i$ (Section~\ref{sec:res-cluster}). ``Align.''\ is the
correlation, across the $60$ actors, between the mean partialled focal covariate of
an actor and its mean log expected count. Both are averaged over senders and
receivers. After either relabelling the correlation lies between $-.01$ and $.01$.
Means over $500$ data sets, evaluated at the true conditional means.}
\end{table}

\paragraph{Design}
The study has six parts, all with $2{,}000$ replications unless stated otherwise.
(A)~Main arms. These hold $\rho = 0.6$ and $\rho_a = 0.30$ in the main condition and cross two nuisance effects, $\beta_2 \in \{0.25, 0.50\}$, with the null
$\beta_1 = 0$ and the alternative $\beta_1 = 0.10$. (B)~A second alternative,
$\beta_1 = 0.03$, in the same arms. With actor effects in the model, $\beta_1 =
0.10$ is detected almost always, so power comparisons between statistics need the
smaller effect. (C)~A grid under the null with $\beta_2 = 0.12$, crossing
$\rho \in \{0, 0.3, 0.6, 0.7, 0.8, 0.9\}$ with $\rho_a \in \{0, 0.15, 0.30\}$ in the
main condition, with $1{,}000$ replications per cell. (D)~Conditions H1 and
H2 under the null in the main arms, $\beta_2 \in \{0.25, 0.50\}$. (E)~Larger networks. The null
simulations of the main arms, and those of conditions H1 and H2 at
$\beta_2 = 0.50$, are repeated with $n = 50$ and with $n = 100$ actors in
each of the three networks, which gives $N = 7{,}350$ and $N = 29{,}700$ dyads. (F)~Actor means. The fits without actor effects of the main condition, H1 and H2
at $\beta_2 = 0.50$ are repeated with the sender mean and the receiver mean of $X_2$,
taken within the network, added to the outcome model and to the residualization.
This part uses the null data sets of parts A, D and E with $n = 20$ and $n = 100$,
and the alternative $\beta_1 = 0.10$ with $n = 20$, for which the data sets of H1 and
H2 are new. Across all
simulated data sets the largest linear predictor is 11.2. With $20$ actors per
network the median of the largest value within a data set is 6.3 under the null at $\beta_2 = 0.50$, 4.2 at $\beta_2 = 0.25$, and between
0.5 and 3.4 in the grid, so the Poisson means need no bound.

\paragraph{Estimation and statistics}
Each data set is fitted twice. The first fit omits the actor effects (``NoRE'') and
has five parameters, $\beta_1$, $\beta_2$ and the three network intercepts. The
second (``RE'') adds one sender and one receiver effect for each of the 60 actors,
penalized by their Gaussian log-density with $\sigma_R$ fixed at its generating
value. At $\rho_a = 0$ the two fits coincide, and only the first is computed. Every
observed and relabelled fit is computed by Newton--Raphson iteration with step
halving to a step tolerance of $10^{-11}$. The model-based variance of
$\hat\beta_1$ is the focal element of the inverse penalized Hessian,
$\mathcal I_{11\cdot2}^{-1}$. HC3 uses the leverages of the full design, that is,
the diagonal of the penalized hat matrix including the network intercepts and, in
the RE fit, the actor effects. The dyadic estimator \eqref{eq:meat} sums score
cross-products over pairs of dyads that share an actor, which here means pairs in the
same network. The DSP residual $\tX_1$ is formed by OLS regression of the focal
covariate on the network indicators and the nuisance covariate, pooled across
networks (Remark~\ref{rem:scope}).

\paragraph{Permutation}
Each permutation draw applies an independent actor relabelling within each network
to rows and columns jointly. The focal array, either raw $X_1$ or the DSP residual
$\tX_1$, is relabelled while the nuisance covariate is held fixed. We use $K = 999$
draws per replication and compute two-sided Monte Carlo $p$-values as
$(1 + \#\{|t^{(k)}| \ge |t^{\mathrm{obs}}|\})/(K+1)$. The observed
labelling counts as one of the $K + 1$ labellings, which keeps the test valid at
every level and the $p$-value above zero \citep{phipson2010}. For the dyadic statistic we
apply the rule of Section~\ref{sec:cluster}. A relabelling whose dyadic variance estimate
is zero or negative (numerically, below $10^{-12}$) is excluded, and the $p$-value
is computed over the $K'$ retained relabellings. When the observed estimate is zero
or negative, the dyadic test is not computed for that data set.

\paragraph{References}
Each statistic is referred to its relabelling distribution and, for comparison, to the
standard normal distribution and to Student-$t$ distributions
with $\nu \in \{18, 55, 1135\}$ degrees of freedom (\ref{app:reference}). These are
the number of dyads minus the number of coefficients ($1{,}140 - 5$), the number of
actors minus the number of coefficients ($60 - 5$), and the number of actors of one
network minus two ($20 - 2$). Cells use
$2{,}000$ replications, except those of the grid, which use $1{,}000$. The $99\%$ Monte
Carlo tolerance band around a nominal $.05$ level is $[.037, .063]$ for $2{,}000$
replications and $[.032, .068]$ for $1{,}000$.

\section{Results}\label{sec:results}

Every rejection rate in this section is for a nominal level of $.05$. Unless stated
otherwise it is computed from the actor-relabelling reference, and next to it we
report the rate obtained when the same observed statistic is referred to a standard
normal distribution. We write $T_W(\hat V)$ for the Wald statistic studentized by
the variance estimator $\hat V$ and name the relabelled array, raw $X_1$ or the DSP
residual $\tX_1$, separately. Rates for the dyadic statistic are over the data sets
in which it is computed. The rule of Section~\ref{sec:cluster} leaves the dyadic test
uncomputed in at most $0.65\%$ of the data sets of any cell, and it excludes on
average at most $0.32\%$ of the relabellings of a data set.

\subsection{The two schemes give the same statistic}\label{sec:res-identity}

Parts A to D of the study consist of $38{,}000$ simulated data sets and $70{,}000$ fitted models,
since the grid at $\rho_a = 0$ uses only the fit without actor effects. Each fitted
model yields three observed statistics, computed once with the raw focal array and once with the DSP residual in the design. Across these $210{,}000$ pairs, the
largest absolute difference between the two versions of a statistic is
$6.2 \times 10^{-8}$. Part E adds $16{,}000$ data sets and $96{,}000$ pairs. In these larger
networks the two versions of the focal coefficient differ by at most
$3.5 \times 10^{-8}$ and the two versions of a statistic by at most
$4.2 \times 10^{-5}$. The latter difference occurs for a model-based statistic of
$82.7$. Part F refits data sets of these parts and adds $4{,}000$ under the alternative.
There the two versions of a statistic differ by at most $1.1 \times 10^{-5}$. In part F the actor means enter the outcome model and the residualization together, as condition (i) of Remark~\ref{rem:scope} requires. Proposition~\ref{prop:fwl} and Corollary~\ref{cor:se}
therefore hold for both fits and all three variance estimators. Every difference
between the two schemes reported below is a difference between their reference
distributions. Because the identity holds only for converged fits, it also
provides a simple check on the fitting routine (Remark~\ref{rem:scope}).

\subsection{When the focal covariate and its residual share their autocorrelation}\label{sec:res-main}\label{sec:res-variance}\label{sec:res-collin}\label{sec:res-beta2}

In the main condition the DSP residual has the same row and column autocorrelation
as the raw focal array, $\rho_V = \rho_X = \rho_a$ (Table~\ref{tab:covstruct}). The
relabelling reference then holds the level of every statistic, for both fits and under either scheme (Table~\ref{tab:main}). In the main arms its rejection
rates lie between $.042$ and $.061$, all inside the tolerance band. The normal
reference fails without actor effects. It rejects a true null at $.737$ and $.898$
with the model-based variance, at $.315$ and $.260$ with HC3, and at $.120$ and
$.167$ with the dyadic variance, for $\beta_2 = 0.25$ and $0.50$ respectively. With
actor effects in the model, the normal reference is close to nominal for the
model-based statistic ($.047$ and $.049$), conservative for HC3 ($.034$ and $.027$)
and liberal for the dyadic statistic ($.086$ and $.108$). The relabelling reference
brings all three to the nominal level.

\begin{table}[tbp]
\centering
\caption{Null rejection at nominal $.05$ in the main arms ($\rho = 0.6$,
$\rho_a = 0.30$, main condition, $2{,}000$ replications per cell).}
\label{tab:main}
\footnotesize
\setlength{\tabcolsep}{4pt}
\begin{tabular}{llcccccc}
\toprule
 & & \multicolumn{3}{c}{$\beta_2 = 0.25$} & \multicolumn{3}{c}{$\beta_2 = 0.50$} \\
\cmidrule(lr){3-5}\cmidrule(lr){6-8}
Actor effects & Statistic & Raw $X$ & DSP & Normal & Raw $X$ & DSP & Normal \\
\midrule
Without & Model-based & .051 & .052 & .737 & .060 & .061 & .898 \\
 & HC3 & .049 & .047 & .315 & .049 & .048 & .260 \\
 & Dyadic & .047 & .048 & .120 & .057 & .059 & .167 \\
\midrule
With & Model-based & .046 & .046 & .047 & .053 & .053 & .049 \\
 & HC3 & .042 & .044 & .034 & .052 & .051 & .027 \\
 & Dyadic & .045 & .045 & .086 & .054 & .055 & .108 \\
\bottomrule
\end{tabular}
\tabnote{``Raw $X$'' and ``DSP'' refer the observed statistic to the relabelling reference
built from the raw focal array ($X$-permutation) or from the DSP residual, and ``Normal'' refers the
same statistic to $\mathcal N(0,1)$. The two schemes give the same observed statistic. The paired exact test of the difference between the two
schemes gives $p > .38$ for every pair. Dyadic rates are over the data sets in which
the test is computed.}
\end{table}

The failures of the normal reference are failures of scale, in the sense of Section~\ref{sec:nonpivot} (Table~\ref{tab:scaleshape}). Without actor
effects the null scale of the model-based statistic is $5.90$ at $\beta_2 = 0.25$
and $13.65$ at $\beta_2 = 0.50$, that of the HC3 statistic $2.03$ and $1.78$, and
that of the dyadic statistic $1.30$ and $1.40$. In every arm the scale-only
prediction lies within $.021$ of the rejection rate under the normal reference, so
the shape of the null distribution adds little. The scale follows from \eqref{eq:uvar}. Without actor effects in the model, the variance of the focal score
is $46$ times its model-based value at $\beta_2 = 0.25$ and $333$ times at
$\beta_2 = 0.50$ (Table~\ref{tab:kappa}). About $70\%$ and $50\%$ of the excess comes
from the second sum of \eqref{eq:uvar}, the covariance between dyads that share an
actor, and the rest from the overdispersion in the first sum. HC3 estimates the first
sum and the model-based variance neither, which is why HC3 fails less badly. The
relabelling reference estimates neither sum. Each relabelled array keeps the actor
structure of the array it is drawn from, so the relabelled designs reproduce each
sum to within $16\%$, and $\kappa$ to within $15\%$ (Table~\ref{tab:kappa}). This is
what makes the permutation test work for network regression. A relabelling leaves
the rows and columns of the relabelled array intact. The row and column
autocorrelation of this array is therefore the same in every draw, which
conditions the reference distribution. However, what changes is which actor
carries which row and column, and with it the expected counts that each row of
focal values is paired with. In particular, under the weak null this matters when
the array is related to the nuisance covariates across actors, which is the case
in Section~\ref{sec:res-struct}.

\begin{table}[tbp]
\centering
\caption{Scale and shape of the null distribution of the observed statistic in the
main arms ($\rho = 0.6$, $\rho_a = 0.30$, $2{,}000$ replications).}
\label{tab:scaleshape}
\footnotesize
\setlength{\tabcolsep}{3pt}
\resizebox{\linewidth}{!}{%
\begin{tabular}{llcccccc}
\toprule
Arm & $\hat V$ & $s$ & Scale-only & Normal & Shape gap & QAP & Normal $-$ QAP \\
\midrule
$\beta_2=.25$, NoRE & model-based & 5.90 & .740 & .737 & $-.003$ & .051 & $.685$ \\
 & HC3 & 2.03 & .335 & .315 & $-.021$ & .049 & $.266$ \\
 & dyadic & 1.30 & .132 & .120 & $-.012$ & .047 & $.073$ \\
\midrule
$\beta_2=.25$, RE & model-based & 0.98 & .045 & .047 & $+.002$ & .046 & $.002$ \\
 & HC3 & 0.89 & .028 & .034 & $+.006$ & .042 & $-.007$ \\
 & dyadic & 1.10 & .076 & .086 & $+.009$ & .045 & $.041$ \\
\midrule
$\beta_2=.50$, NoRE & model-based & 13.65 & .886 & .898 & $+.012$ & .060 & $.839$ \\
 & HC3 & 1.78 & .271 & .260 & $-.011$ & .049 & $.211$ \\
 & dyadic & 1.40 & .161 & .167 & $+.006$ & .057 & $.110$ \\
\midrule
$\beta_2=.50$, RE & model-based & 1.03 & .058 & .049 & $-.009$ & .053 & $-.004$ \\
 & HC3 & 0.88 & .027 & .027 & $.000$ & .052 & $-.025$ \\
 & dyadic & 1.16 & .091 & .108 & $+.017$ & .054 & $.054$ \\
\bottomrule
\end{tabular}%
}
\tabnote{The statistic is the same under both schemes. The null scale $s$ is
$\operatorname{median}|T| / \Phi^{-1}(.75)$, ``Scale-only'' is $2\Phi(-1.96/s)$,
``Normal'' refers $T$ to $\mathcal N(0,1)$, and ``Shape gap'' is their difference.
``QAP'' is the rejection rate under the relabelling reference built from the raw
focal array, and ``Normal $-$ QAP'' is the difference between the two rates (negative where
the normal reference is conservative).
Differences are computed before rounding. For the dyadic statistic the replications
are those in which it is computed.}
\end{table}

\begin{figure}[t]
\centering
\includegraphics[width=\textwidth]{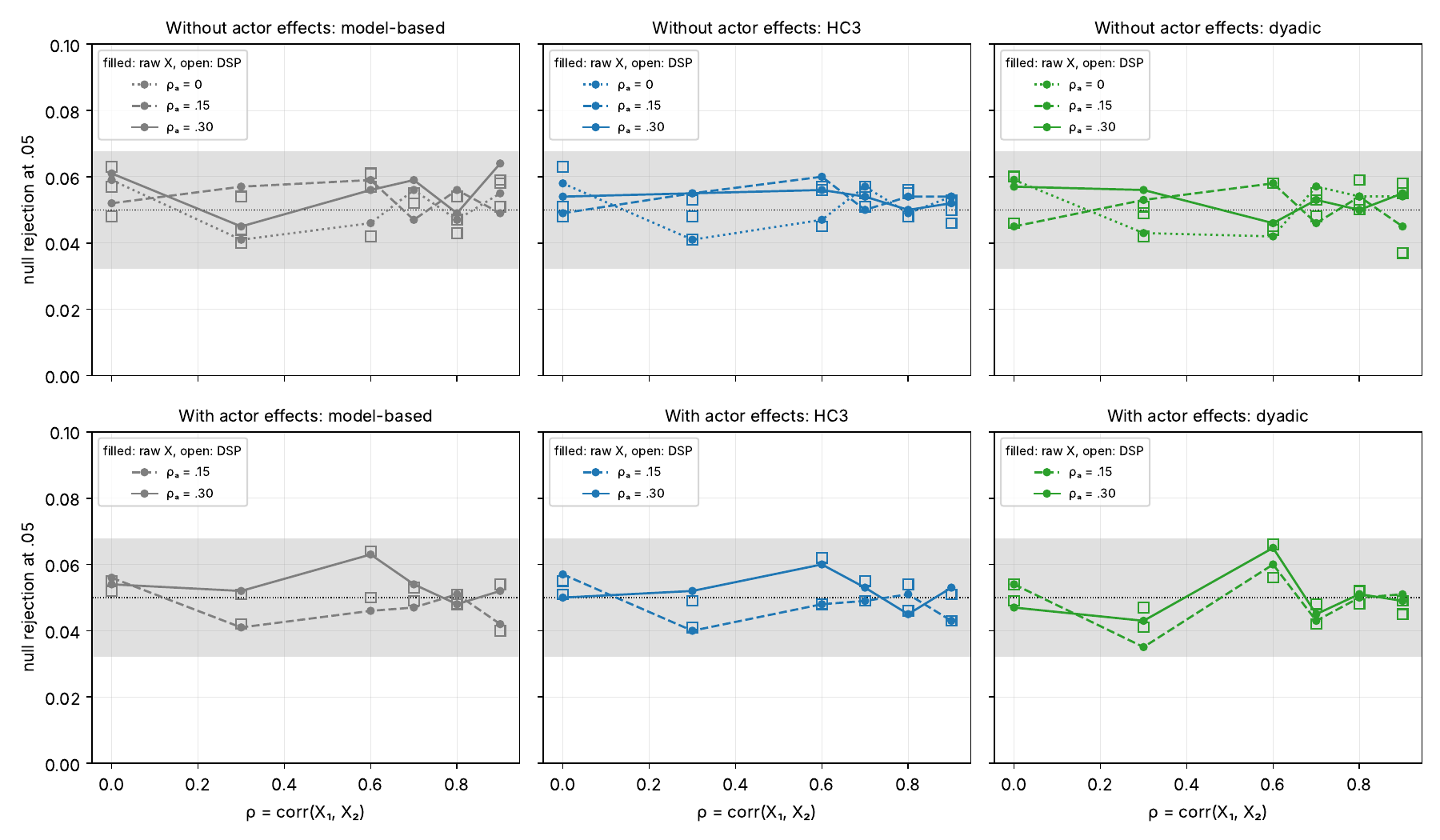}
\caption{Null rejection at nominal $.05$ under the relabelling reference in the grid
($\beta_2 = 0.12$, main condition, $1{,}000$ replications per point),
against the correlation $\rho$ between the covariates, without (top) and with (bottom)
actor effects in the model. Line style marks $\rho_a$. Filled markers relabel the raw
focal array and open markers the DSP residual. The grey band is the $99\%$ Monte Carlo
tolerance band around $.05$. At $\rho_a = 0$ the two fits coincide, and only the fit
without actor effects is shown. Rates for the dyadic statistic are over the data sets
in which it is computed.}
\label{fig:collin}
\end{figure}

The grid at $\beta_2 = 0.12$ shows the same pattern across collinearity and
autocorrelation (Figure~\ref{fig:collin}, with the rates in Table~\ref{tab:grid} of
\ref{app:grid}). Under the relabelling reference its $180$ rejection rates lie
between $.035$ and $.066$, with mean $.051$, and all fall inside the band for
$1{,}000$ replications. They show no trend in $\rho$. At $\rho = 0.9$ every statistic
is as close to the nominal level as at $\rho = 0$, under both fits. The normal
reference again depends on the actor effects (Table~\ref{tab:gridnormal}). Without
actor effects in the model it
is close to nominal for the model-based and HC3 statistics at $\rho_a = 0$ ($.041$ to
$.062$). Once the data contain actor effects it fails. At $\rho_a = 0.15$ the
model-based statistic then rejects at $.17$ to $.22$ and HC3 at $.12$ to $.14$, and at
$\rho_a = 0.30$ at $.62$ to $.66$ and $.31$ to $.33$. With actor effects in the model
these two statistics stay between $.034$ and $.058$ under the normal reference. The
dyadic statistic is liberal under the normal reference throughout, at $.075$ to
$.123$, including at $\rho_a = 0$.

Whether the raw focal array or its DSP residual is relabelled makes no difference
to the rejection rates in the main condition. Under the null the two rates differ by
at most $.002$ in the main arms and by at most $.008$ in the grid, where $2$ of the
$90$ paired exact tests fall below $.05$, both for the dyadic statistic (marked in
Table~\ref{tab:grid}). This is what
the design implies. When the raw focal array and its residual have the same
autocorrelation, the two relabellings reproduce the same structure.

\subsection{When the collinearity runs through parts with different autocorrelation}\label{sec:res-struct}\label{sec:res-scheme}\label{sec:res-reference}

In conditions H1 and H2 the residual has a different autocorrelation from the raw
focal array (Table~\ref{tab:covstruct}), and the two schemes separate
(Table~\ref{tab:struct}). Without actor effects, $X$-permutation is conservative in H1 and liberal in H2.
This is the direction expected in Section~\ref{sec:design}. In H1 the model-based statistic rejects at
$.011$ and $.027$ and HC3 at $.008$ and $.020$, for $\beta_2 = 0.25$ and $0.50$. In H2
the model-based statistic rejects at $.155$ at both values and HC3 at $.152$ and
$.144$. DSP instead gives $.053$ to $.064$ in H2 and $.072$
to $.105$ in H1. In each of these eight comparisons the paired exact test gives
$p < .001$. Under the normal reference all three statistics fail without actor
effects, as in the main condition, at $.65$ to $.90$ for the model-based statistic,
$.17$ to $.38$ for HC3 and $.13$ to $.19$ for the dyadic statistic.

\begin{table}[tbp]
\centering
\caption{Null rejection at nominal $.05$ in conditions H1 and H2 ($\rho = 0.6$,
$\rho_a = 0.30$, $2{,}000$ replications per cell), with columns as in
Table~\ref{tab:main}.}
\label{tab:struct}
\footnotesize
\setlength{\tabcolsep}{4pt}
\begin{tabular}{llcccccc}
\toprule
 & & \multicolumn{3}{c}{$\beta_2 = 0.25$} & \multicolumn{3}{c}{$\beta_2 = 0.50$} \\
\cmidrule(lr){3-5}\cmidrule(lr){6-8}
Actor effects & Statistic & Raw $X$ & DSP & Normal & Raw $X$ & DSP & Normal \\
\midrule
\multicolumn{8}{l}{\emph{H1, actor-level collinearity}} \\
Without & Model-based & .011 & .082 & .648 & .027 & .105 & .866 \\
 & HC3 & .008 & .072 & .173 & .020 & .094 & .174 \\
 & Dyadic & .060 & .056 & .148 & .077 & .070 & .191 \\
\addlinespace
With & Model-based & .048 & .050 & .054 & .050 & .048 & .051 \\
 & HC3 & .047 & .051 & .037 & .051 & .049 & .027 \\
 & Dyadic & .051 & .054 & .094 & .048 & .048 & .091 \\
\midrule
\multicolumn{8}{l}{\emph{H2, dyad-level collinearity}} \\
Without & Model-based & .155 & .061 & .767 & .155 & .064 & .897 \\
 & HC3 & .152 & .053 & .377 & .144 & .058 & .302 \\
 & Dyadic & .061 & .065 & .131 & .067 & .069 & .180 \\
\addlinespace
With & Model-based & .052 & .053 & .052 & .054 & .051 & .051 \\
 & HC3 & .070 & .062 & .054 & .057 & .054 & .030 \\
 & Dyadic & .063 & .058 & .117 & .059 & .058 & .101 \\
\bottomrule
\end{tabular}
\tabnote{In H1 the part common to the two covariates is actor-level and
the specific parts are dyad-level, in H2 the reverse (Table~\ref{tab:covstruct}). The
paired exact test of the difference between the schemes gives $p < .001$ for every
model-based and HC3 pair without actor effects and for HC3 in H2 at
$\beta_2 = 0.25$ with actor effects, $p$ between $.02$ and $.05$ for three dyadic
pairs, and $p > .09$ elsewhere. Dyadic rates are over the data sets in which the test
is computed.}
\end{table}

Table~\ref{tab:kappa} accounts for the direction of these errors. In H1 the raw focal
array is more autocorrelated than its residual ($\rho_X = 0.27$ against
$\rho_V = 0.07$). Its relabellings produce designs with a larger score variance than
the observed design, by $66\%$ at $\beta_2 = 0.25$ and $17\%$ at $\beta_2 = 0.50$, so
the reference is too wide and the test conservative. In H2 the raw focal array is less autocorrelated than its residual ($0.18$ against $0.38$). Its relabellings understate
the score variance by $45\%$ and $40\%$, and the test is liberal. The DSP relabellings
keep the autocorrelation of the residual, the part of the focal covariate that is left
once the nuisance covariate has been partialled out, and they stay within $27\%$ of
the observed $\kappa$. In H2 this brings the test close to the nominal level. In H1
they understate $\kappa$ by $19\%$ and $27\%$, and the test rejects at $.072$ to
$.105$. The split of $\kappa$ in Table~\ref{tab:kappa} locates the shortfall. The DSP
relabellings reproduce the overdispersion part as well as in the main condition, but
only $66\%$ and $37\%$ of the part that comes from dyads sharing an actor. In H1 the
actor components of the residual and of the nuisance covariate are the same variable,
the common sender and receiver parts of \eqref{eq:xgen}. Across the actors of a data
set, the mean of the partialled focal covariate then correlates $0.57$ with the mean
of the log expected count (at $\beta_2 = 0.50$, Table~\ref{tab:kappa}). A relabelling
keeps the autocorrelation of the residual but moves its values to other actors, and
the correlation falls to zero. In H2 the same correlation is $-0.38$, and the DSP
relabellings still reproduce $94\%$ and $86\%$ of the actor part. In the main
condition it is zero.

The dyadic statistic and the fits with actor effects depend much less on the
structure. With actor effects every statistic rejects at $.047$ to $.063$ under both
schemes, with one exception. HC3 in H2 at $\beta_2 = 0.25$ rejects at $.070$ under $X$-permutation and at $.062$ under DSP (paired $p < .001$). Without actor effects
the dyadic statistic rejects at $.056$ to $.065$ at $\beta_2 = 0.25$ and at $.067$ to
$.077$ at $\beta_2 = 0.50$, under either scheme. This excess appears only in H1 and
H2, the conditions in which the residual is aligned with the expected counts at the
actor level, and we examine it in Section~\ref{sec:res-cluster}.

These conditions give the recommendation to use DSP a substantive basis. When a variance estimator misses part of the dependence, the null
distribution of the statistic is governed by the autocorrelation of the focal
covariate after the nuisance covariates have been partialled out. The DSP residual
carries that autocorrelation, and the raw focal array in general does not. Where the two
coincide, as in the main condition, the schemes agree. Where they differ, $X$-permutation rejects at up to three times the nominal level in H2 and at a fraction of it in H1, while DSP is close to nominal in H2 and moderately liberal
in H1.
Section~\ref{sec:res-largern} shows how these errors change with the size of the
network.

\subsection{The dyadic statistic and the actor effects}\label{sec:res-cluster}

The dyadic variance is the only one of the three that targets both sums of
\eqref{eq:uvar}, and the dyadic statistic is the least sensitive to the structure of
the collinearity. Under the normal reference it is nevertheless liberal in every cell
of the study, at $.075$ to $.095$ even in the grid at $\rho_a = 0$
(Table~\ref{tab:gridnormal}), where dyads that share an actor are independent. As expected in Section~\ref{sec:design}, the reason is the small number of actors. Let $v_i$
be the variance, given the covariates, of the part of the focal score that runs
through sender $i$, and take $(\sum_i v_i)^2 / \sum_i v_i^2$ as the effective number
of senders that carry the score, and likewise for receivers. At the true means and
error covariance \eqref{eq:ecov}, the average of the two is about $9$ of the $60$
actors at $\beta_2 = 0.50$, in the main condition as well as in H1 and H2, and between
$17$ and $19$ at $\beta_2 = 0.25$ (Table~\ref{tab:kappa}). With so few effective units
the dyadic variance is
noisy and falls short of the variance of $\hat\beta_1$. The null scale of the dyadic
statistic is $1.30$ and $1.40$ without actor effects and $1.10$ and $1.16$ with them
(Table~\ref{tab:scaleshape}).

To see how the relabelling reference handles this, we refitted the $2{,}000$ null
data sets of the main condition, H1 and H2 at $\beta_2 = 0.50$ without actor effects.
We drew $199$ new relabellings of each array and stored every relabelled estimate and
dyadic variance. In all three conditions the dyadic variance of the observed fits falls short of the
variance of $\hat\beta_1$ across data sets, by a factor of $1.79$ in the main
condition, $1.99$ in H1 and $1.77$ in H2 (Table~\ref{tab:dyaddecomp}). Within each
reference the relabelled fits
fall short by a similar factor, between $1.54$ and $1.65$. The relabelling reference
therefore carries most of the shortfall with it, which is why the test is close to
nominal although the dyadic statistic is not pivotal. The match is not exact. In this
run the test rejects at $.057$ and $.060$ in the main condition and at $.061$ to
$.080$ in H1 and H2 ($X$-permutation and DSP), close to the rates of
Tables~\ref{tab:main} and~\ref{tab:struct}. The small excess appears only in the two conditions in which
the residual is aligned with the expected counts across actors, a feature that no relabelling keeps. We have not isolated
how the alignment produces it. Neither the largest linear predictor of a data set nor
the concentration of its score on a few actors predicts which data sets reject.

\begin{table}[tbp]
\centering
\caption{Supplementary run for the dyadic statistic without actor effects at
$\beta_2 = 0.50$ ($\rho = 0.6$, $\rho_a = 0.30$).}
\label{tab:dyaddecomp}
\footnotesize
\setlength{\tabcolsep}{4pt}
\begin{tabular}{lrccccc}
\toprule
 & & \multicolumn{3}{c}{$\operatorname{Var}(\hat\beta_1) / \operatorname{E}(\hat V)$} & \multicolumn{2}{c}{Rejection} \\
\cmidrule(lr){3-5}\cmidrule(lr){6-7}
Condition & Data sets & Observed & Raw $X$ reference & DSP reference & Raw $X$ & DSP \\
\midrule
Main & 1{,}999 & 1.79 & 1.65 & 1.64 & .057 & .060 \\
H1 & 1{,}987 & 1.99 & 1.64 & 1.56 & .080 & .071 \\
H2 & 1{,}999 & 1.77 & 1.54 & 1.57 & .061 & .069 \\
\bottomrule
\end{tabular}
\tabnote{The null data sets of Tables~\ref{tab:main} and~\ref{tab:struct} are refitted with $199$ new relabellings of each array. ``Observed'' is the variance of $\hat\beta_1$ across data sets divided by
the mean dyadic variance $\hat V$. The reference columns give the same ratio within
the relabelled fits of a data set, averaged over data sets. ``Rejection'' is at
nominal $.05$ under the reference of $199$ relabellings. Data sets in which the test
is not computed are left out.}
\end{table}

With actor effects in the model the dyadic statistic holds its level in every cell of
the study, at $.035$ to $.066$ under either scheme, and at $.048$ to $.063$ in H1 and
H2. Without them it is the most robust of the three statistics, but with $20$ actors
per network ($60$ actors in total) it can exceed the nominal level by about half when the focal covariate and its residual differ in autocorrelation.
Section~\ref{sec:res-largern} shows how the shortfall of the dyadic variance
changes when the networks have more actors.

\subsection{Power}\label{sec:res-power}

Because every statistic holds its level in the main arms, Table~\ref{tab:power}
reports rejection rates under the alternatives without a size correction. With actor
effects in the model every statistic detects $\beta_1 = 0.10$ in at least $99.9\%$ of
the data sets. Without them, power against the same effect is $.30$ to $.32$ at
$\beta_2 = 0.25$ and $.23$ to $.26$ at $\beta_2 = 0.50$. Against $\beta_1 = 0.03$ the
fits without actor effects reject at $.065$ to $.086$, barely above their size, while
the fits with actor effects reach $.31$ to $.38$ at $\beta_2 = 0.25$ and $.80$ to $.87$
at $\beta_2 = 0.50$. The loss reflects the inflation of the score variance in
Table~\ref{tab:kappa}, which the reference reproduces under the null and which an
effect has to overcome under the alternative. A larger nuisance effect
raises the means and with them the information about $\beta_1$, which is why power
rises with $\beta_2$ when the actor effects are modelled. Without them the inflation
rises faster, from $46$ to $333$, and power falls. Leaving the actor effects out keeps the
permutation test valid and costs most of its power.

Among the fits with actor effects the model-based statistic is the most powerful and
the dyadic statistic the least. At $\beta_1 = 0.03$ the dyadic statistic reaches
$.314$ and $.799$, against $.378$ and $.870$ for the model-based statistic, which is
$83\%$ and $92\%$ of its power. The two schemes give the same power. The largest
difference between them is $.010$ (paired $p = .013$), for the dyadic statistic
without actor effects at $\beta_2 = 0.50$ and $\beta_1 = 0.10$, where DSP rejects
slightly more often. With $24$ comparisons in the table, one paired $p$-value of this
size is unremarkable.

\begin{table}[tbp]
\centering
\caption{Rejection at nominal $.05$ under the alternatives in the main arms
($\rho = 0.6$, $\rho_a = 0.30$, $2{,}000$ replications per cell), relabelling
reference, as raw $X$ / DSP.}
\label{tab:power}
\footnotesize
\setlength{\tabcolsep}{4pt}
\begin{tabular}{llcccc}
\toprule
 & & \multicolumn{2}{c}{$\beta_1 = 0.10$} & \multicolumn{2}{c}{$\beta_1 = 0.03$} \\
\cmidrule(lr){3-4}\cmidrule(lr){5-6}
Actor effects & Statistic & $\beta_2 = 0.25$ & $\beta_2 = 0.50$ & $\beta_2 = 0.25$ & $\beta_2 = 0.50$ \\
\midrule
Without & Model-based & .321 / .319 & .254 / .256 & .077 / .079 & .083 / .086 \\
 & HC3 & .301 / .300 & .234 / .236 & .067 / .070 & .065 / .066 \\
 & Dyadic & .295 / .298 & .226 / .236 & .072 / .072 & .065 / .065 \\
\midrule
With & Model-based & 1.000 / 1.000 & 1.000 / 1.000 & .378 / .377 & .870 / .866 \\
 & HC3 & 1.000 / 1.000 & 1.000 / 1.000 & .355 / .357 & .862 / .859 \\
 & Dyadic & .999 / .999 & 1.000 / 1.000 & .314 / .320 & .799 / .804 \\
\bottomrule
\end{tabular}
\tabnote{No size correction is applied, because every test holds
its level in these arms (Table~\ref{tab:main}). The paired exact test of the
difference between the schemes gives $p = .013$ for the dyadic statistic without
actor effects at $\beta_1 = 0.10$ and $\beta_2 = 0.50$, and $p > .08$ for every other
pair. Dyadic rates are over the data sets in which the test is computed.}
\end{table}

\subsection{Larger networks}\label{sec:res-largern}

All results so far are for networks of $20$ actors. To see whether they depend on
this size, we repeated the null simulations with $50$ and with $100$ actors per
network, for the main condition at both nuisance effects
(Table~\ref{tab:largern}) and for conditions H1 and H2 at $\beta_2 = 0.50$
(Table~\ref{tab:largernstruct}). The three networks, both fits and both relabelled
arrays were kept, with $2{,}000$ data sets per cell and $K = 999$. The alternatives
were not repeated at these sizes.

The relabelling test keeps its level when the networks grow. With actor effects in
the model, the rejection rates of all three statistics lie between $.043$ and
$.060$ at both sizes. Without actor effects, the HC3 and dyadic statistics lie
between $.046$ and $.061$. The model-based statistic without actor effects is the
exception. At $\beta_2 = 0.25$ it rejects at $.052$ to $.056$. At $\beta_2 = 0.50$ it rejects at $.069$ under $X$-permutation and at $.066$ under DSP with $50$ actors, and at $.065$ and $.066$ with $100$. With $20$ actors these rates were $.060$ and $.061$. These four rates are
just above the tolerance band. As in the networks of $20$ actors, the two
relabelled arrays give the same test, and the paired exact test gives $p > .14$
for every pair.

The normal reference changes with the size of the network, and the direction
depends on the variance estimator. The model-based and HC3 variances ignore the
dependence between dyads that share an actor, and the number of such pairs
grows with the number of actors. Hence, without actor effects their null scale
rises. At $\beta_2 = 0.50$ the null scale of the model-based statistic goes from $13.65$ with $20$ actors to $34.77$ with $100$, and the normal reference then rejects a true null at $.960$. For HC3 the scale goes from $1.78$ to $2.95$ and the rejection rate reaches $.510$. The dyadic variance targets this dependence and is estimated from more actors in a larger network. Hence its null scale falls. Without actor effects it goes from $1.40$ to $1.13$ at $\beta_2 = 0.50$, and the rejection rate under the normal reference from $.167$ to $.077$. With actor effects the scale goes from $1.16$ to $1.02$ and the rejection rate from $.108$ to $.062$. At $\beta_2 = 0.25$ the rejection rate with $100$ actors is $.057$ without actor effects and $.054$ with them. No observed dyadic
estimate was zero or negative with $50$ or $100$ actors, and on average fewer
than $0.001\%$ of the relabellings of a data set were excluded.

The shortfall of the dyadic variance therefore shrinks when the number of actors
grows, in line with the linear-model results of \citet{marrs2023}. It has not
disappeared at $100$ actors. The relabelling reference is still needed at these
sizes, also for the dyadic statistic.

In conditions H1 and H2 the errors of Section~\ref{sec:res-struct} do not shrink with
the size of the network. Without actor effects they grow
(Table~\ref{tab:largernstruct}). $X$-permutation becomes more conservative in H1. There the model-based statistic rejects at $.027$, $.018$ and $.014$ with $20$, $50$ and $100$ actors. In H2 it becomes more liberal, and the same statistic rejects at $.155$, $.181$ and $.202$. HC3 follows the same course. DSP stays close to its rates at $20$ actors in H2, between $.055$ and $.073$.
In H1 it does not. Under DSP the model-based statistic rejects at $.105$, $.128$ and $.152$ with $20$, $50$ and $100$ actors, and HC3 at $.094$, $.102$ and $.121$. The dyadic statistic changes least. It
rejects at $.060$ to $.082$ under either scheme and at every size, so its excess
neither grows much nor disappears.

With actor effects in the model none of this remains. Every statistic rejects at
$.040$ to $.059$ in H1 and H2, at all three sizes and under either scheme.

The size of the network matters in the way expected in
Section~\ref{sec:design}. The second sum of \eqref{eq:uvar} grows faster than the
first and dominates in a large network. An actor attribute shows this most simply. If the partialled
focal covariate were an attribute $a_i$ of the sender, the focal score would be the
sum over senders of $a_i$ times the sum of the errors in row $i$. With the sender
effects left in the errors, the variance of such a row sum grows with $n^2$. Both errors sit in this second sum. The raw focal array has the wrong autocorrelation there, and in H1 the DSP residual loses its alignment with
the expected counts across actors when it is relabelled. In H2 the DSP relabellings reproduce most of the
actor part already with $20$ actors, and the test stays close to the nominal
level.

This reading can be tested. The residual is correlated with $X_2$ across actors
because the residualization uses one slope where the actor-level and the dyad-level
relation between the covariates differ (Section~\ref{sec:design}). Adding the sender mean
and the receiver mean of $X_2$ to the outcome model, and with it to the
residualization, gives the two levels their own slopes. We refitted the null data
sets of the main condition, H1 and H2 at $\beta_2 = 0.50$ in this way, without actor
effects, with $20$ and with $100$ actors (Table~\ref{tab:actormeans}). The alignment
of the partialled focal covariate with the expected counts disappears. In H1 it
falls from $.59$ to $-.05$ with $20$ actors and from $.85$ to $-.03$ with $100$, and
in H2 from $-.37$ and $-.46$ to $.04$ and $.01$. DSP then holds its level. All
eighteen of its rejection rates lie between $.040$ and $.063$, inside the tolerance
band, where the fits without the actor means gave up to $.152$ in H1 with $100$
actors. The main condition is unchanged. $X$-permutation is not repaired. With the
model-based and HC3 statistics it rejects at $.000$ to $.008$ in H1 and at $.137$ to
$.164$ in H2, because the autocorrelation of the raw focal array still differs from
that of the residual. Only the dyadic statistic holds its level under
$X$-permutation, at $.050$ to $.059$.

The actor means repair the level and not the power. Against $\beta_1 = 0.10$ in the
main condition with $20$ actors, DSP rejects at $.233$ to $.264$ with the actor means
and at $.236$ to $.256$ without them, against at least $.999$ when the actor effects
are modelled (Table~\ref{tab:power}). In H1 and H2 the fits without the actor means
were not run under the alternative, so those rates of Table~\ref{tab:actormeans}
have no counterpart. The actor means also have a cost for the dyadic statistic. In
H1 with $20$ actors the test is not computed in $28$ of the $2{,}000$ data sets,
against $13$ without the actor means.

\begin{table}[tbp]
\centering
\caption{Null rejection at nominal $.05$ in the main condition
($\rho = 0.6$, $\rho_a = 0.30$) for three networks of $n = 20$, $50$ and $100$
actors, $2{,}000$ replications per cell.}
\label{tab:largern}
\footnotesize
\setlength{\tabcolsep}{2pt}
\begin{tabular}{llccccccccc}
\toprule
Actor& & \multicolumn{3}{c}{$n = 20$} & \multicolumn{3}{c}{$n = 50$} & \multicolumn{3}{c}{$n = 100$} \\
\cmidrule(lr){3-5}\cmidrule(lr){6-8}\cmidrule(lr){9-11}
effects & Statistic & Raw $X$ & DSP & Normal & Raw $X$ & DSP & Normal & Raw $X$ & DSP & Normal \\
\midrule
\multicolumn{11}{l}{\emph{$\beta_2 = 0.25$}} \\
Without & Model-based & .051 & .052 & .737 & .054 & .052 & .832 & .055 & .056 & .880 \\
 & HC3 & .049 & .047 & .315 & .047 & .049 & .502 & .049 & .049 & .614 \\
 & Dyadic & .047 & .048 & .120 & .048 & .049 & .076 & .047 & .050 & .057 \\
\addlinespace
With & Model-based & .046 & .046 & .047 & .052 & .053 & .053 & .049 & .049 & .048 \\
 & HC3 & .042 & .044 & .034 & .052 & .052 & .042 & .046 & .048 & .043 \\
 & Dyadic & .045 & .045 & .086 & .057 & .060 & .082 & .046 & .046 & .054 \\
\midrule
\multicolumn{11}{l}{\emph{$\beta_2 = 0.50$}} \\
Without & Model-based & .060 & .061 & .898 & .069 & .066 & .934 & .065 & .066 & .960 \\
 & HC3 & .049 & .048 & .260 & .055 & .054 & .408 & .046 & .047 & .510 \\
 & Dyadic & .057 & .059 & .167 & .061 & .060 & .105 & .053 & .051 & .077 \\
\addlinespace
With & Model-based & .053 & .053 & .049 & .048 & .046 & .047 & .050 & .051 & .049 \\
 & HC3 & .052 & .051 & .027 & .045 & .043 & .032 & .049 & .049 & .041 \\
 & Dyadic & .054 & .055 & .108 & .048 & .047 & .062 & .050 & .054 & .062 \\
\bottomrule
\end{tabular}
\tabnote{Columns as in Table~\ref{tab:main}, whose
rates are repeated under $n = 20$. At $n = 50$ and $n = 100$ the paired exact test
of the difference between the two schemes gives $p > .14$ for every pair, and the
dyadic test is computed in every data set.}
\end{table}

\begin{table}[tbp]
\centering
\caption{Null rejection at nominal $.05$ in conditions H1 and H2 at
$\beta_2 = 0.50$ ($\rho = 0.6$, $\rho_a = 0.30$) for three networks of $n = 20$,
$50$ and $100$ actors, $2{,}000$ replications per cell.}
\label{tab:largernstruct}
\footnotesize
\setlength{\tabcolsep}{2pt}
\begin{tabular}{llccccccccc}
\toprule
Actor & & \multicolumn{3}{c}{$n = 20$} & \multicolumn{3}{c}{$n = 50$} & \multicolumn{3}{c}{$n = 100$} \\
\cmidrule(lr){3-5}\cmidrule(lr){6-8}\cmidrule(lr){9-11}
effects & Statistic & Raw $X$ & DSP & Normal & Raw $X$ & DSP & Normal & Raw $X$ & DSP & Normal \\
\midrule
\multicolumn{11}{l}{\emph{H1, actor-level collinearity}} \\
Without & Model-based & .027 & .105 & .866 & .018 & .128 & .917 & .014 & .152 & .936 \\
 & HC3 & .020 & .094 & .174 & .010 & .102 & .257 & .008 & .121 & .361 \\
 & Dyadic & .077 & .070 & .191 & .077 & .072 & .123 & .081 & .082 & .106 \\
\addlinespace
With & Model-based & .050 & .048 & .051 & .047 & .047 & .049 & .050 & .049 & .051 \\
 & HC3 & .051 & .049 & .027 & .047 & .043 & .033 & .049 & .050 & .040 \\
 & Dyadic & .048 & .048 & .091 & .048 & .050 & .066 & .055 & .053 & .065 \\
\midrule
\multicolumn{11}{l}{\emph{H2, dyad-level collinearity}} \\
Without & Model-based & .155 & .064 & .897 & .181 & .071 & .939 & .202 & .073 & .957 \\
 & HC3 & .144 & .058 & .302 & .154 & .057 & .433 & .166 & .055 & .530 \\
 & Dyadic & .067 & .069 & .180 & .060 & .060 & .098 & .065 & .068 & .088 \\
\addlinespace
With & Model-based & .054 & .051 & .051 & .051 & .051 & .049 & .044 & .042 & .042 \\
 & HC3 & .057 & .054 & .030 & .049 & .048 & .036 & .043 & .040 & .030 \\
 & Dyadic & .059 & .058 & .101 & .055 & .050 & .071 & .044 & .041 & .053 \\
\bottomrule
\end{tabular}
\tabnote{Columns as in
Table~\ref{tab:main}. The rates under $n = 20$ repeat those of
Table~\ref{tab:struct}. The paired exact test of the difference between the two
schemes gives $p < .001$ for every model-based and HC3 pair without actor effects
and $p > .04$ for every other pair. Dyadic rates are over the data sets in which the
test is computed.}
\end{table}

\begin{table}[tbp]
\centering
\caption{Fits without actor effects, without and with the sender mean and the
receiver mean of $X_2$ in the outcome model and in the residualization
($\rho = 0.6$, $\rho_a = 0.30$, $\beta_2 = 0.50$, $2{,}000$ replications per cell,
the same data sets in both rows of a condition).}
\label{tab:actormeans}
\footnotesize
\setlength{\tabcolsep}{4pt}
\begin{tabular}{llrccc}
\toprule
Condition & Actor means & Align. & Model-based & HC3 & Dyadic \\
\midrule
\multicolumn{6}{l}{\emph{Null, $n = 20$}} \\
Main & Without & $.00$ & .060 / .061 & .049 / .048 & .057 / .059 \\
 & With & $.00$ & .060 / .059 & .049 / .048 & .055 / .060 \\
\addlinespace
H1 & Without & $.59$ & .027 / .105 & .020 / .094 & .077 / .070 \\
 & With & $-.05$ & .008 / .062 & .004 / .058 & .059 / .051 \\
\addlinespace
H2 & Without & $-.37$ & .155 / .064 & .144 / .058 & .067 / .069 \\
 & With & $.04$ & .137 / .058 & .146 / .044 & .052 / .057 \\
\midrule
\multicolumn{6}{l}{\emph{Null, $n = 100$}} \\
Main & Without & $.00$ & .065 / .066 & .046 / .047 & .053 / .051 \\
 & With & $.00$ & .064 / .063 & .048 / .048 & .050 / .052 \\
\addlinespace
H1 & Without & $.85$ & .014 / .152 & .008 / .121 & .081 / .082 \\
 & With & $-.03$ & .000 / .060 & .000 / .051 & .052 / .048 \\
\addlinespace
H2 & Without & $-.46$ & .202 / .073 & .166 / .055 & .065 / .068 \\
 & With & $.01$ & .164 / .043 & .157 / .040 & .058 / .058 \\
\midrule
\multicolumn{6}{l}{\emph{Alternative $\beta_1 = 0.10$, $n = 20$}} \\
Main & Without & $.00$ & .254 / .256 & .234 / .236 & .226 / .236 \\
 & With & $.00$ & .260 / .264 & .240 / .238 & .232 / .233 \\
\addlinespace
H1 & Without & $.59$ & -- & -- & -- \\
 & With & $-.05$ & .186 / .415 & .132 / .406 & .367 / .345 \\
\addlinespace
H2 & Without & $-.37$ & -- & -- & -- \\
 & With & $.04$ & .280 / .173 & .282 / .160 & .150 / .154 \\
\bottomrule
\end{tabular}
\tabnote{Rejection at nominal $.05$ under
the relabelling reference, as $X$-permutation / DSP. ``Align.''\ is the actor-level
alignment of Table~\ref{tab:kappa}, computed on the data sets of this table.
Without the actor means, the null rates repeat those of Tables~\ref{tab:main},
\ref{tab:struct}, \ref{tab:largern} and~\ref{tab:largernstruct} and the rates under
the alternative those of Table~\ref{tab:power}. H1 and H2 were not run under the
alternative without the actor means. Dyadic rates are over the data sets in which
the test is computed.}
\end{table}

\section{Empirical illustration}\label{sec:example}

We illustrate the argument on the Freeman EIES network data
\citep{freeman1979,freeman1980,wasserman1994}. The network records $n = 32$
researchers, and the outcome $Y_{ij}$ is the count of electronic messages sent
from researcher $i$ to researcher $j$, giving $N = 992$ directed dyads once
the diagonal is excluded. The message counts are heavily overdispersed, with
an off-diagonal maximum of $559$ and a mean of $15.2$, and they show
pronounced sender and receiver heterogeneity. The focal covariate is the
acquaintanceship rating of $j$ reported by $i$ at the start of the study on a
scale from $0$ to $4$. The nuisance covariates are a same-discipline
indicator, which distinguishes the $17$ sociologists from the $15$ other
researchers, and the absolute difference in log citations. The focal covariate
is moderately correlated with both nuisance variables, with raw correlations
of $r = 0.20$ and $r = -0.25$.

We estimate two specifications. Specification~A includes an intercept and the
three dyadic covariates, while Specification~B adds sender and receiver fixed
effects. In both specifications, DSP residualizes the focal covariate on the
intercept and the two dyadic nuisance covariates by OLS. Because the removed
component lies in the column span of both outcome designs,
Proposition~\ref{prop:fwl} applies to both specifications, and on these data
the focal coefficients and standard errors of the raw-$X$ and DSP fits agree
exactly. Because the observed statistics are common to both schemes,
Table~\ref{tab:eies} reports them once per specification, while the QAP
$p$-values are reported for both $X$-permutation and DSP using $K = 9{,}999$
actor relabellings. We also report four diagnostics, the identity check of Proposition~\ref{prop:fwl}, the $\hat\mu$-weighted variance inflation factor, the empirical
permutation scale $\sigma_\pi$, and the number of relabellings excluded because
their dyadic variance estimate was zero or negative.

The two checks of Section~\ref{sec:disc} can be read from the covariates. The
sample autocorrelation \eqref{eq:rhohat} of the acquaintanceship ratings is
$.10$ over rows and $.19$ over columns, and that of their DSP residual is $.11$
and $.18$. Residualization therefore leaves the autocorrelation of the focal
array as it is, and $X$-permutation and DSP should give nearly the same test.
Table~\ref{tab:eies} shows that they do. Across the $32$ actors, the mean of the
DSP residual correlates $-.01$ with the mean of the same-discipline indicator,
for senders and for receivers alike. With the mean of the citation difference it
correlates $.25$ for senders and $-.11$ for receivers. The sender value is not
negligible, although it is well below the $.59$ of condition H1
(Table~\ref{tab:actormeans}). Specification~B models the actor effects, which
removes the concern there.

\begin{table}[tbp]
\centering
\caption{EIES messages: the acquaintanceship effect under three variance
estimators.}
\label{tab:eies}
\footnotesize
\setlength{\tabcolsep}{4pt}
\begin{tabular}{llccccccr}
\toprule
Spec & $\hat V$ & $\hat\beta_1$ & s.e. & $t$ & $p$(QAP-$X$) & $p$(QAP-DSP) & $\sigma_\pi$ & \multicolumn{1}{c}{$p(t_\nu)$} \\
\midrule
A & dyadic      & 0.538 & 0.114 & 4.72 & .0243 & .0209 & 3.52 & $2.7{\times}10^{-6}$ \\
  & HC3         & 0.538 & 0.087 & 6.17 & .0090 & .0071 & 2.35 & $1.0{\times}10^{-9}$ \\
  & model-based & 0.538 & 0.007 & 74.1 & .0003 & .0003 & 23.8 & $<\!10^{-15}$ \\
\midrule
B & dyadic      & 0.319 & 0.070 & 4.53 & .0276 & .0304 & 2.29 & $6.8{\times}10^{-6}$ \\
  & HC3         & 0.319 & 0.073 & 4.37 & .0006 & .0007 & 1.19 & $1.4{\times}10^{-5}$ \\
  & model-based & 0.319 & 0.010 & 32.7 & .0001 & .0001 & 7.14 & $<\!10^{-15}$ \\
\bottomrule
\end{tabular}
\tabnote{$p$(QAP-$X$) and $p$(QAP-DSP) are the relabelling $p$-values
across $K = 9{,}999$ permutations. Because the two schemes share the observed
statistic exactly (Proposition~\ref{prop:fwl}), it is reported once.
$\sigma_\pi$ is the standard deviation of the statistic across relabellings
under $X$-permutation (for the dyadic statistic, across the relabellings
retained under the rule of Section~\ref{sec:cluster}), the empirical null scale that
a parametric reference assumes equals one. The dyadic row is the recommended statistic. For comparison,
$p(t_\nu)$ uses the dyad-based degrees of freedom ($\nu = 988$ and $926$), and
actor-based bounds move it by less than $.04$ (\ref{app:reference}).}
\end{table}

The dyadic cluster-robust Wald statistic gives a clear and moderate answer.
Acquaintanceship predicts message volume conditional on disciplinary and
citation similarity, with $\hat\beta_1 = 0.319$ once actor fixed effects are
included (Specification~B), and the relabelling test on
$T_W(\hat V_{\mathrm{dyad}})$ returns $p = .0276$ under $X$-permutation and
$p = .0304$ under DSP. Equivalently, $2.8\%$ of actor relabellings produce a
test statistic at least as large as the observed $|t| = 4.53$. The finding is
supported at the conventional $0.05$ level across both permutation schemes.
Specification~B models the actor heterogeneity, here with fixed effects. In
our simulations the dyadic test held its level in every condition once the
actor effects were modelled (there as random effects).

The diagnostics show why the dyadic-robust row is preferred. The $\sigma_\pi$
column reports the null scale of each statistic under relabelling, which a
parametric reference assumes to be one. For the model-based statistic,
$\sigma_\pi$ is $23.8$ in Specification~A and $7.14$ in Specification~B,
reflecting severe overdispersion with deviance-to-df ratios of $42.0$ and
$8.2$, and its $t$-values of $74.1$ and $32.7$ reflect a standard error that
is an order of magnitude too small. For the dyadic statistic, $\sigma_\pi$
lies between $2.29$ and $3.52$ under $X$-permutation and between $2.02$ and $2.56$
under DSP, showing that even the cluster-robust statistic
departs substantially from a standard normal distribution. This scale
distortion explains why the last column of Table~\ref{tab:eies} reports a
parametric $p$-value of $6.8 \times 10^{-6}$ where the relabelling test
reports $.0276$. The gap measures how badly a fixed parametric table misstates
the null distribution when relabelled acquaintanceship arrays retain chance
alignment with the strong actor heterogeneity in message counts. Furthermore,
the $\hat\mu$-weighted variance inflation factor rises from $1.06$ in
Specification~A to between $1.7$ and $2.0$ in Specification~B as the actor
fixed effects reweight the design, illustrating the mechanism of
Section~\ref{sec:nonpivot} at moderate levels. An applied researcher relying on the
parametric $p$-value would claim overwhelming statistical evidence where the
relabelling test on the dyadic statistic indicates moderate evidence at the
$0.05$ level. The HC3 and model-based relabelling tests return smaller
$p$-values. In our simulations these tests also stayed close to their level
once the actor effects were modelled, but the simulated outcomes carried no
dependence beyond the actor effects, such as reciprocity. The dyadic variance
is the only one of the three that targets every dependence between dyads that
share an actor, reciprocal pairs included, and we base the conclusion on it.

The dyadic variance estimate is zero or negative for some relabellings in the
EIES data as well, even at $n = 32$ actors. Of the $9{,}999$ relabellings,
$136$ under $X$-permutation and $88$ under DSP gave a zero or negative estimate
in Specification~A, and $436$ and $399$ in Specification~B once the actor fixed
effects are estimated. Following the rule of Section~\ref{sec:cluster}, these
relabellings are excluded, and the dyadic $p$-values in Table~\ref{tab:eies}
are computed over the remaining ones, between $95.6\%$ and $99.1\%$ of the
draws. The observed estimates are positive in both specifications, so the test
is computed. An analyst inspecting only the observed standard error would see
nothing unusual, which is why the number of excluded relabellings belongs in
the reported diagnostics. Replication code and the exact
Newton--Raphson IRLS fitter, verified against standard GLM software, are \ifanonymous provided in the supplementary materials\else available from the authors and will be deposited in a public repository on publication\fi.

\section{Discussion}\label{sec:disc}

Table~\ref{tab:controlmap} brings the results together. It lists the sources of
extra variance in the focal score and what each part of the procedure does with
them. Two sources are present whenever actors differ in activity or popularity,
the overdispersion of the counts and the covariance between dyads that share an
actor. The model-based variance estimates neither, HC3 estimates the first, and
the dyadic variance estimates both but noisily. The relabelling reference
estimates nothing. It reproduces whatever row and column structure the
relabelled array carries, with the outcome and the nuisance covariates held
fixed, and the choice between $X$-permutation and DSP decides whether that structure is the one the
statistic depends on. Including the actor effects in the outcome model removes
the shared-actor dependence from the errors. When the collinearity runs through
the actor components, the residual is also aligned with the expected counts
across actors. No relabelling keeps that alignment, and only the actor effects
in the model remove it.

\begin{table}[tbp]
\centering
\caption{What controls each source of extra variance in the focal score
\eqref{eq:uvar}, summarizing Section~\ref{sec:results}.}
\label{tab:controlmap}
\scriptsize
\setlength{\tabcolsep}{2.5pt}
\resizebox{\linewidth}{!}{%
\begin{tabular}{>{\raggedright\arraybackslash}p{2.6cm}ccccccc}
\toprule
 & \multicolumn{3}{c}{Variance estimator} & \multicolumn{2}{c}{Added to the outcome model} & \multicolumn{2}{c}{Relabelled array} \\
\cmidrule(lr){2-4}\cmidrule(lr){5-6}\cmidrule(lr){7-8}
Source & \shortstack{Model-\\based} & HC3 & Dyadic & \shortstack{Actor\\effects} & \shortstack{Actor means of\\nuisance covariates} & Raw $X_1$ & \shortstack{DSP\\residual} \\
\midrule
Collinearity alone & no effect & no effect & no effect & no effect & no effect & no effect & no effect \\[3pt]
Overdispersion (dyad-level sum) & misses & estimates & estimates & removes & \shortstack{does not\\remove} & reproduces & reproduces \\[3pt]
Shared-actor dependence (actor-level sum) & misses & misses & \shortstack{estimates,\\noisily} & removes & \shortstack{does not\\remove} & \shortstack{only if\\$\rho_X = \rho_V$} & reproduces \\[3pt]
Residual autocorrelation differs, $\rho_V \ne \rho_X$ (H1, H2) & sensitive & sensitive & insensitive & insensitive & \shortstack{does not\\remove} & \shortstack{does not\\reproduce} & reproduces \\[3pt]
Residual aligned with the means across actors (H1, H2) & sensitive & sensitive & slightly & removes & \shortstack{removes, DSP\\then holds its level} & \shortstack{does not\\reproduce} & \shortstack{does not\\reproduce} \\[3pt]
Few actors, $9$ to $19$ effective with $n = 20$ & \shortstack{not\\affected} & conservative & biased low & \shortstack{does not\\fix} & \shortstack{does not\\fix} & \shortstack{mostly\\reproduces} & \shortstack{mostly\\reproduces} \\
\bottomrule
\end{tabular}%
}
\tabnote{The first three columns
are variance estimators. The next two are additions to the outcome model, the
sender and receiver effects or the sender and receiver means of the nuisance
covariates (Section~\ref{sec:res-largern}). The last two are the relabelling reference
built from the raw focal array or from the DSP residual. ``Estimates'' and
``misses'' refer to the variance estimate, ``removes'' to the source itself, and
``reproduces'' to the relabelled designs. For the two sources that arise only in
conditions H1 and H2, an entry for a variance estimator says whether the
relabelling test built on it is sensitive to the source.}
\end{table}

\paragraph{Recommendations}
We suggest six practices for Poisson MRQAP with correlated dyadic covariates.
Each addresses one of the sources in Table~\ref{tab:controlmap}.

\begin{enumerate}
\item \textbf{Refer a studentized statistic, rather than a raw coefficient, to
  its QAP permutation distribution.} The null distribution of a raw coefficient
  shifts with the collinearity of the covariates and with the size of the
  nuisance effect. A parametric reference assumes that dyads are independent.
  In our simulations it rejected a true null far too often in designs where the
  relabelling reference held the nominal level.
\item \textbf{Residualize by OLS on the fixed-effect nuisance design of the
  outcome model, and relabel the DSP residual.} The raw and the residualized fit
  must then return the same focal coefficient and standard error, which is a
  check on the implementation (Remark~\ref{rem:scope}; the \texttt{MrQAP}
  package of \citealt{krause2026mrqap} residualizes this way by default).
  Relabelling the DSP residual builds the reference distribution for the same
  partialled focal covariate as the observed fit. When the focal covariate and
  its residual differed in autocorrelation, $X$-permutation was conservative or
  liberal, and more so in larger networks. DSP stayed close to the nominal level
  when the collinearity ran through the dyad-level parts. Without actor effects
  it was liberal when the collinearity ran through the actor-level parts.
\item \textbf{Model the sender and receiver heterogeneity in the outcome
  model.} Without actor effects the relabelling test keeps its level in the main
  condition but has little power, because the dependence between dyads that
  share an actor inflates the variance of the focal score. With sender and
  receiver effects the power was close to one for the same focal effect, and
  every statistic rejected a true null at $.035$ to $.070$ in every condition
  and network size we examined. The relabelling reference remains available for
  this specification, the additive part of the models of
  \citet{hoff2005,hoff2021}, because the actor effects travel with the labels.
\item \textbf{Use the two-way dyadic cluster-robust variance, and state how
  zero or negative variance estimates were handled.} The dyadic variance
  targets both the overdispersion and the dependence between dyads that share
  an actor, and the relabelling test built on it was the least sensitive of the
  three to how the collinearity is structured. With the actor effects modelled
  it costs some power relative to the model-based statistic. Its estimate can be
  zero or negative when actors are few. Exclude such relabellings from the
  reference distribution, report their number, and do not compute the test when
  the observed estimate is zero or negative. The same rule is applied to the
  observed fit and to every relabelled fit, so the retained statistics remain
  exchangeable under the null. Hence, a large share of excluded relabellings
  does not by itself invalidate the test. The share depends on the data. In our
  design it rises with the size of the largest means and is unchanged by
  modelling the actor effects. It was at most $0.32\%$ on average in our
  simulations and up to $4.4\%$ in the EIES data, whose counts are far
  heavier-tailed. When the tested quantity combines several coefficients, the
  dyadic covariance matrix of these coefficients can be replaced by its positive
  semidefinite part \citep{cameron2011}, in the observed fit and in every
  relabelling alike, so that no relabelling is excluded. We have not evaluated
  this repair in our simulations.
\item \textbf{Report $\sigma_\pi$, the standard deviation of the studentized
  statistic across relabellings.} A parametric $p$-value assumes that the
  studentized statistic has unit scale under the null. In the EIES data its
  scale under relabelling lies between $1.19$ and $23.8$. A value far from one
  tells the reader that an asymptotic $p$-value for that statistic is
  unreliable, at no computational cost beyond the permutations.
\item \textbf{When the actor effects cannot be modelled, add the sender mean and
  the receiver mean of the nuisance covariates to the model, and report two
  checks.} Without actor effects two things can go wrong. Both can be seen in the
  covariates before any model is fitted. First, the raw focal array and its DSP
  residual can differ in their row and column autocorrelation
  \eqref{eq:rhohat}, in which case $X$-permutation can be far off in either
  direction. Second, the DSP residual can still be related to a nuisance
  covariate across actors. The residual is uncorrelated with the nuisance
  covariates over the dyads. However, when both are averaged over the dyads of
  each actor, these actor means can be correlated. A relabelling gives the rows
  and columns of the residual to other actors, so this relation is lost in the
  reference distribution. Adding the sender mean and the receiver mean of the
  nuisance covariates to the outcome model, and with it to the residualization,
  removes the correlation. DSP then held its level in our simulations, with $20$
  and with $100$ actors (Section~\ref{sec:res-largern}). $X$-permutation was not
  repaired, and the power remained that of a fit without actor effects. We
  tested this with one nuisance covariate only. We recommend reporting the two
  autocorrelations and the correlation between the actor means, so that readers
  can see whether these terms mattered.
\end{enumerate}

\paragraph{Weighted residualization}
For weighted linear regression with known weights, \citet{terbraak2022}
establishes that permuting the unweighted predictor residual directly is the
most robust option across weight configurations, whereas standardizing the
residual by the square root of the weights is preferable only when the error
variance is inversely proportional to those weights. The Poisson fit sits
between these two cases. Its working IRLS regression has error variance
inversely proportional to the GLM weight, which is the configuration where
standardization can improve power, but the weights $\hat\mu$ are estimated
rather than fixed and change with every relabelled fit. We have used
unweighted OLS residualization throughout on the strength of the
robustness result of \citet{terbraak2022}, and Proposition~\ref{prop:fwl} holds for that form. Whether a 
$\hat\mu$-weighted residualization would improve power in
generalized linear models, and at what cost to robustness when the mean is
misspecified, remains an open question. The score-based construction of
\citet{hemerik2020}, which orthogonalizes in the metric of the information
matrix for independent observations, is the natural point of comparison.

\paragraph{Limitations and extensions}
Most of our simulations use small networks, with $n = 20$ actors in each of
$G = 3$ networks, in order to expose finite-sample behaviour. The null
simulations of the main condition and of conditions H1 and H2 were repeated
with $50$ and $100$ actors, the alternatives were not. Its dependence
runs entirely through sender and receiver effects. The covariates are generated
independently of these effects, the outcome has no dyad-level error term and no
reciprocity, and the random-effects fits use the generating variance of the
actor effects. Network outcomes such as the EIES messages can carry dependence
of other kinds, and of our three variance estimators only the dyadic one
targets it. Without actor effects in the model, the dyadic test ran slightly
above the nominal level in the conditions in which the focal covariate and its
residual differ in autocorrelation, at $.056$ to $.077$ with $20$ actors and at
$.060$ to $.082$ with $50$ and $100$. It does not shrink in larger networks. It disappears when the sender mean and the
receiver mean of the nuisance covariate are added to the model, which points to the alignment of the residual with the
expected counts across actors as its source. We tested these actor means with one
nuisance covariate and in these three conditions only.

Our networks are also of equal size, generated with the same parameters, and
disjoint in their actors. The relabelling within networks, with the same
permutation applied to rows and columns, is the largest group that maps the
set of dyads onto itself, and the test is valid as a stratified permutation
test when actors are exchangeable within a network under the null, the
coefficients are common across networks, and no actor belongs to more than
one network. Exchanging whole networks of equal size would enlarge the group
by at most $G!$ and would require the networks to be exchangeable in the
relabelled array, which fails as soon as they differ in the distribution of
the covariates. Unequal sizes leave the argument intact and change the
weighting. A network contributes $n_g(n_g - 1)$ dyads to the score and the
actor-level part of $\kappa$ grows with $n_g - 2$, so the pooled statistic
and the dyadic variance are governed by the largest networks, and the
reference reproduces each network's dependence separately. A single larger
network raises $\kappa$ for the same reason and gives the dyadic estimator
more actors to work with. Networks that share actors, as in multiplex or
repeated observations, require one permutation applied to every network and
a variance that clusters across networks on the same actor, and lie outside
our results.

Our covariates are also light-tailed, Poisson with mean four, and the largest
linear predictor in the study is $11.2$. Tails matter for the first sum of
\eqref{eq:uvar}. A relabelling moves the focal values to other dyads while the
means stay where they are, so the reference reproduces that sum only insofar as
large squared values of the partialled focal covariate do not sit at the dyads
with the largest means. In our design they do so mildly. At the $1\%$ of dyads
with the largest means the squared partialled focal covariate is typically
$1.3$ times its average at $\beta_2 = 0.25$ and $1.2$ times at
$\beta_2 = 0.50$, against $0.8$ to $0.9$ after relabelling. The relabelled
designs reproduce $91\%$ and $84\%$ of the overdispersion part of $\kappa$
(Table~\ref{tab:kappa}). Covariates with heavier tails that share their
extremes would strengthen this alignment, because the dyads with the largest
means would then also carry the most variable focal values. The alignment
concerns individual dyads rather than rows and columns, so no relabelling of
either array keeps it, and our study does not include such covariates. Two
measures limit the risk. Modelling the actor effects removes the overdispersion
they create, which in our design is all of it, and a transformation such as
$\log(1 + x)$ shortens the tails of a count covariate where its interpretation
allows. Extensions to heavy-tailed covariates, to negative-binomial and binary
links, to unequal and larger actor counts, to estimated variance components, and
to the finite-sample corrections of \citet{tabordmeehan2019} within a
permutation framework are natural next steps.

\section{Conclusion}\label{sec:conc}

We studied MRQAP permutation tests for Poisson regression on network count
data. A studentized statistic referred to its relabelling distribution held the
nominal level for the model-based, HC3 and dyadic cluster-robust variances
whenever the focal covariate and its residual shared their autocorrelation, where
a normal reference did not. When the two autocorrelations differed, the choice
between $X$-permutation and DSP mattered, the dyadic statistic was the least
affected, and the errors of the fits without actor effects grew with the size of
the network. With sender and receiver effects in the model every statistic
stayed close to its level in every condition, and the power was close to one
where it had been about $.3$ without them.

All simulations used three separate networks, with the actors relabelled
independently within each network and one model fitted to the pooled dyads.
Hence, every result above was obtained under the second extension that
\citet{shi2025} left open, and the relabelling test works in that setting. The
EIES example applies the same procedure to a single network.

Researchers should model the sender and receiver effects, studentize the focal
coefficient with the dyadic cluster-robust variance, evaluate it against the relabelling distribution of the DSP residual, and report
$\sigma_\pi$, the number of excluded relabellings and the autocorrelation of
the focal covariate before and after residualization. Readers can then see how
close to pivotal the statistic was, and which reference it needed.

\section*{Declaration of generative AI and AI-assisted technologies in the manuscript preparation process}
During the preparation of this work the authors used Claude (Anthropic) to assist with
writing and checking the simulation and analysis code, checking numbers in the text against
the simulation outputs, and drafting and editing parts of the text. After using this tool,
the authors reviewed and edited the content as needed and take full responsibility for the
content of the published article.

\section*{Data availability}
The code that reruns the simulations and rebuilds every table and number, and the
simulation outputs, are \ifanonymous provided as supplementary material\else
available from the authors\fi{} and will be deposited in a public repository on
publication. The EIES data are publicly available.

\clearpage
\appendix
% Tables, figures and equations are numbered per appendix (Table A.1, B.1, ...).
\makeatletter
\@addtoreset{table}{section}
\@addtoreset{figure}{section}
\@addtoreset{equation}{section}
\makeatother

\section{Proofs}\label{app:proofs}

\begin{proof}[Proof of Proposition~\ref{prop:fwl}]
The OLS residual \eqref{eq:resid} satisfies $\tX_1 = X_1 - \bZ \bm a$ with
$\bm a = (\bZ\trans\bZ)^{-1}\bZ\trans X_1$. Hence the column spaces coincide,
with $\operatorname{col}([\tX_1,\bZ]) = \operatorname{col}([X_1,\bZ]) =: \mathcal C$. The Poisson log-likelihood $\ell(\bm\eta) = \sum_\ell \{Y_\ell \eta_\ell - e^{\eta_\ell}\}$ is strictly concave in $\bm\eta$ and is
maximized over the same linear subspace $\mathcal C$ by both fits, so the
maximizer $\hat{\bm\eta}$ is unique and common to both. Expressing
$\hat{\bm\eta}$ in the two full-rank bases gives
$\hat{\bm\eta} = \hat\beta_1^{X} X_1 + \bZ \hat{\bm c}_1 = \hat\beta_1^{\mathrm{DSP}} \tX_1 + \bZ \hat{\bm c}_2 = \hat\beta_1^{\mathrm{DSP}} X_1 + \bZ(\hat{\bm c}_2 - \hat\beta_1^{\mathrm{DSP}}\bm a)$,
and matching the unique coordinate on $X_1$ yields
$\hat\beta_1^{X} = \hat\beta_1^{\mathrm{DSP}}$.
\end{proof}

\begin{proof}[Proof of Corollary~\ref{cor:se}]
Write the two design matrices as $\bm X = [X_1, \bZ]$ and
$\tilde{\bm X} = [\tX_1, \bZ] = \bm X\bm T$ with
\begin{equation}
\bm T = \begin{pmatrix} 1 & \bm 0\trans \\ -\bm a & \bI \end{pmatrix}, \qquad \bm T^{-1} = \begin{pmatrix} 1 & \bm 0\trans \\ \bm a & \bI \end{pmatrix}.
\end{equation}
By Proposition~\ref{prop:fwl}, the two fits share $\hat{\bm\mu}$ and hence
$\bW$, the residuals $Y - \hat{\bm\mu}$, and the leverages, which depend on
$\operatorname{col}(\bm X)$ alone. Every variance estimator in this paper has
the sandwich form $\bm A^{-1}\bm B\bm A^{-1}$ with $\bm A = \bm X\trans\bW\bm X$
and $\bm B = \bm X\trans\bm\Omega\bm X$, where $\bm\Omega$ is an $N \times N$
matrix built from $\hat{\bm\mu}$, the residuals, and the leverages. Setting
$\bm\Omega = \bW$ gives the model-based variance, a leverage-adjusted diagonal
$\bm\Omega$ gives HC3, and the actor-sharing pattern of \eqref{eq:meat} gives
the dyadic cluster-robust estimator. Since $\bm\Omega$ is common to both fits,
the substitution $\bm X \mapsto \bm X\bm T$ transforms the sandwich covariance
matrix into $\bm T^{-1}\bm A^{-1}\bm B\bm A^{-1}\bm T^{-\!\top}$. Because the
first row of $\bm T^{-1}$ is $(1, \bm 0\trans)$, the $(1,1)$ element, which is
the focal variance, remains unchanged. The identity therefore holds for all
three standard errors, and consequently for all three studentized statistics.
\end{proof}

\paragraph{Penalized fits with actor effects}
Both proofs carry over to the fit with sender and receiver effects. Write its
linear predictor as $\bm\eta = \bm X\bm\theta + \bm U\bm b$, where $\bm U$
holds the actor indicators and $\bm b$ the actor effects, and its objective as
$\ell(\bm\eta) - \tfrac12\bm b\trans\bm\Lambda\bm b$ with a fixed positive
definite $\bm\Lambda$. The reparameterization $\bm X \mapsto \bm X\bm T$
changes only $\bm\theta$, which carries no penalty, so the two fits maximize
the same objective over the same set of $(\bm X\bm\theta, \bm b)$ and share
its unique maximizer. The penalized Hessian $\bm H = [\bm X, \bm U]\trans\bW[\bm
X, \bm U] + \operatorname{diag}(\bm 0, \bm\Lambda)$ becomes $\bm T_*\trans\bm
H\bm T_*$ with $\bm T_* = \operatorname{diag}(\bm T, \bI)$, whose inverse has
first row $(1, \bm 0\trans)$. The leverages, the partialled focal covariate and
the dyadic score contributions depend only on $\operatorname{col}([\bm X, \bm
U])$ and on the penalty, so the focal element of each sandwich is unchanged.

\section{Asymptotic references across degrees of freedom}\label{app:reference}

Section~\ref{sec:reference} argues that no Student-$t$ reference distribution is well
defined for a dyadic Wald statistic, because the exchangeable unit is the actor rather
than the dyad and the degrees of freedom are not identified.
Table~\ref{tab:appref} refers the same fitted statistics of the main arms to
Student-$t$ distributions at the dyad-based choice $\nu = 1135$ and at the actor-based
choices $\nu = 55$ and $\nu = 18$, alongside the relabelling reference. Without actor
effects no choice of $\nu$ comes near the nominal level. The rates lie between $.099$
and $.898$, and moving from $1135$ to $18$ degrees of freedom lowers them by at most
$.031$. With actor effects the model-based statistic rejects at $.034$ to $.049$, HC3
is conservative and becomes more so as $\nu$ falls ($.019$ to $.034$), and the dyadic
statistic is liberal under every choice ($.068$ to $.107$). The relabelling reference
is the only one that brings all three statistics to the nominal level under both fits.

\begin{table}[h]
\centering
\caption{Null rejection at nominal $.05$ by reference distribution in the main arms
($\rho = 0.6$, $\rho_a = 0.30$, $2{,}000$ replications).}
\label{tab:appref}
\footnotesize
\setlength{\tabcolsep}{4pt}
\begin{tabular}{llcccc}
\toprule
Arm & $\hat V$ & QAP & $t_{1135}$ & $t_{55}$ & $t_{18}$ \\
\midrule
$\beta_2=.25$, NoRE & model-based & .051 & .737 & .733 & .726 \\
 & HC3 & .049 & .314 & .304 & .283 \\
 & dyadic & .047 & .120 & .110 & .099 \\
\midrule
$\beta_2=.25$, RE & model-based & .046 & .047 & .042 & .034 \\
 & HC3 & .042 & .034 & .030 & .024 \\
 & dyadic & .045 & .086 & .081 & .068 \\
\midrule
$\beta_2=.50$, NoRE & model-based & .060 & .898 & .894 & .890 \\
 & HC3 & .049 & .259 & .253 & .228 \\
 & dyadic & .057 & .167 & .160 & .145 \\
\midrule
$\beta_2=.50$, RE & model-based & .053 & .049 & .045 & .037 \\
 & HC3 & .052 & .027 & .025 & .019 \\
 & dyadic & .054 & .107 & .098 & .084 \\
\bottomrule
\end{tabular}
\tabnote{``QAP'' is the relabelling
reference built from $X$-permutation. For the dyadic statistic the rates are over
the data sets in which it is computed.}
\end{table}

\section{Rejection rates in the collinearity grid}\label{app:grid}

Table~\ref{tab:grid} lists the rejection rates plotted in Figure~\ref{fig:collin}, and
Table~\ref{tab:gridnormal} the rates of the same statistics under the normal reference.

\begin{table}[!ht]
\centering
\caption{Null rejection at nominal $.05$ under the relabelling reference in the grid
($\beta_2 = 0.12$, main condition, $1{,}000$ replications per cell), as raw
$X$ / DSP.}
\label{tab:grid}
\scriptsize
\setlength{\tabcolsep}{3pt}
\begin{tabular}{lllcccccc}
\toprule
$\rho_a$ & Actor effects & Statistic & $\rho = 0$ & $0.3$ & $0.6$ & $0.7$ & $0.8$ & $0.9$ \\
\midrule
0 & Without & Model-based & .059/.063 & .041/.040 & .046/.042 & .056/.052 & .047/.047 & .055/.059 \\
 &  & HC3 & .058/.063 & .041/.041 & .047/.045 & .057/.055 & .049/.048 & .054/.053 \\
 &  & Dyadic & .059/.060 & .043/.042 & .042/.044 & .057/.055 & .054/.050 & .054/.058 \\
\midrule
.15 & Without & Model-based & .052/.048 & .057/.054 & .059/.061 & .047/.052 & .056/.054 & .049/.051 \\
 &  & HC3 & .049/.048 & .055/.053 & .060/.057 & .050/.057 & .054/.056 & .054/.050 \\
 &  & Dyadic & .045/.046 & .053/.051 & .058/.058 & .046/.048 & .054/.059 & .045/.037\rlap{$^{\dagger}$} \\
 & With & Model-based & .056/.055 & .041/.042 & .046/.050 & .047/.049 & .051/.051 & .042/.040 \\
 &  & HC3 & .057/.055 & .040/.041 & .048/.048 & .049/.049 & .051/.054 & .043/.043 \\
 &  & Dyadic & .054/.054 & .035/.041 & .060/.056 & .043/.042 & .050/.052 & .051/.049 \\
\midrule
.30 & Without & Model-based & .061/.057 & .045/.044 & .056/.057 & .059/.055 & .049/.043 & .064/.058 \\
 &  & HC3 & .054/.051 & .055/.048 & .056/.056 & .054/.051 & .050/.055 & .052/.046 \\
 &  & Dyadic & .057/.060 & .056/.049\rlap{$^{\dagger}$} & .046/.045 & .053/.053 & .050/.052 & .055/.055 \\
 & With & Model-based & .054/.052 & .052/.051 & .063/.064 & .054/.053 & .048/.048 & .052/.054 \\
 &  & HC3 & .050/.051 & .052/.049 & .060/.062 & .053/.055 & .045/.046 & .053/.051 \\
 &  & Dyadic & .047/.049 & .043/.047 & .065/.066 & .045/.048 & .051/.048 & .049/.045 \\
\bottomrule
\end{tabular}
\tabnote{At $\rho_a = 0$ the two fits coincide, and only the fit without actor
effects is computed. The $99\%$ Monte Carlo tolerance band is $[.032, .068]$. A dagger
marks the two pairs in which the paired exact test of the difference between the
schemes gives $p < .05$ (both $p = .039$). Dyadic rates are over the data sets in
which the test is computed.}
\end{table}

\begin{table}[!ht]
\centering
\caption{Null rejection at nominal $.05$ under the normal reference in the grid
($\beta_2 = 0.12$, main condition, $1{,}000$ replications per cell).}
\label{tab:gridnormal}
\scriptsize
\setlength{\tabcolsep}{3pt}
\begin{tabular}{lllcccccc}
\toprule
$\rho_a$ & Actor effects & Statistic & $\rho = 0$ & $0.3$ & $0.6$ & $0.7$ & $0.8$ & $0.9$ \\
\midrule
0 & Without & Model-based & .062 & .041 & .042 & .054 & .043 & .057 \\
 &  & HC3 & .062 & .041 & .048 & .058 & .048 & .053 \\
 &  & Dyadic & .095 & .075 & .087 & .090 & .083 & .092 \\
\midrule
.15 & Without & Model-based & .173 & .222 & .211 & .186 & .209 & .195 \\
 &  & HC3 & .119 & .131 & .137 & .119 & .129 & .126 \\
 &  & Dyadic & .088 & .087 & .105 & .086 & .097 & .083 \\
 & With & Model-based & .055 & .041 & .045 & .045 & .051 & .043 \\
 &  & HC3 & .056 & .039 & .048 & .049 & .050 & .046 \\
 &  & Dyadic & .112 & .083 & .100 & .097 & .097 & .101 \\
\midrule
.30 & Without & Model-based & .640 & .633 & .653 & .621 & .662 & .616 \\
 &  & HC3 & .325 & .320 & .332 & .312 & .328 & .316 \\
 &  & Dyadic & .119 & .116 & .111 & .116 & .112 & .123 \\
 & With & Model-based & .049 & .048 & .058 & .056 & .047 & .056 \\
 &  & HC3 & .042 & .039 & .046 & .049 & .034 & .039 \\
 &  & Dyadic & .103 & .088 & .116 & .103 & .102 & .099 \\
\bottomrule
\end{tabular}
\tabnote{The observed statistic is the same under both schemes. At $\rho_a = 0$ the two fits
coincide, and only the fit without actor effects is computed. Dyadic rates are over
the data sets in which the test is computed.}
\end{table}

\clearpage
\bibliographystyle{elsarticle-harv}
\bibliography{refs}

\end{document}